\documentclass{IEEEtran}
\usepackage{cite}
\usepackage{amsmath,amssymb,amsfonts}
\usepackage{algorithmic}
\usepackage{graphicx}
\usepackage{textcomp}
\usepackage{physics}
\usepackage{siunitx}
\usepackage{bm}
\usepackage{esint}

\usepackage{xcolor}
\usepackage[caption=false,font=footnotesize]{subfig}
\usepackage{float}
\usepackage{url}
\usepackage{enumitem} 
\usepackage[frak=esstix]{mathalpha}
\AtBeginDocument{\RenewCommandCopy\qty\SI}
\def\BibTeX{{\rm B\kern-.05em{\sc i\kern-.025em b}\kern-.08em
    T\kern-.1667em\lower.7ex\hbox{E}\kern-.125emX}}
\usepackage{soul}
\usepackage[table]{xcolor}
\usepackage{hyperref}

\newcommand*\diff{\mathop{}\!\mathrm{d}}
\newcommand{\rvec}{\bm{r}}
\newcommand{\rpvec}{\bm{r}'}
\newcommand{\rfvec}{\bm{r}_\mathrm{f}}

\newcommand{\mvec}{\bm{m}}
\newcommand{\pvec}{\bm{p}}
\newcommand{\E}{\bm{E}}
 
\newcommand{\Hvec}{\bm{H}}
 
\newcommand{\EAM}{E_\mathrm{AM}}
\newcommand{\EAMmax}{E_\mathrm{AM}^\mathrm{max}}
\newcommand{\xhat}{\hat{\bm{x}}}
\newcommand{\yhat}{\hat{\bm{y}}}
\newcommand{\zhat}{\hat{\bm{z}}}
\newcommand{\mhat}{\hat{\bm{m}}}
\newcommand{\phat}{\hat{\bm{p}}} 
\newcommand{\zetavec}{\boldsymbol{\zeta}}

\newcommand{\JTI}{\mathcal{J}_\mathrm{TI}}
\newcommand{\SAR}{\mathrm{SAR}_\text{10g}}
\newcommand{\Vfoc}{\mathcal{V}_{50\%}}
\newcommand{\e}{\mathrm{e}}

\newcommand{\imj}{\mathrm{j}}

\begin{document}
\title{Microwave Focusing with Temporal Interference for Non-Invasive Deep Brain Stimulation}
\author{Mika Söderström, Melker Carlsson, Patrik Nicolausson, and Mariana Dalarsson

\thanks{This work was supported by the Swedish Research Council, project number 2023-04887, and by the Göran Gustafsson foundation, grant number 2406.}
\thanks{M. S., M. C., P. N., and M. D. are with the Department of Electromagnetics and Plasma Physics, KTH Royal Institute of Technology, 100 44 Stockholm, Sweden (e-mail: \href{mikaso@kth.se}{mikaso@kth.se}).}}

\maketitle

\begin{abstract}
Deep Brain Stimulation (DBS) is an effective treatment for neurological disorders but requires invasive surgery. This work presents a method for non-invasive DBS, based on microwave focusing of amplitude-modulated electric fields using an external antenna array of magnetic point dipoles. The proposed method combines iterative time reversal (iTR) and temporal interference (TI) optimization to jointly address electromagnetic field focusing and physiologically relevant neural stimulation. Antenna element positions, orientations, frequencies, amplitudes, and phases are optimized to localize stimulation within a target region. The method is evaluated in an anatomically realistic voxel head model with heterogeneous and lossy tissue properties. Systematic numerical studies, including perturbation analysis and statistical evaluation, demonstrate consistent spatial localization and robustness across all reported configurations. Safety is quantified using specific absorption rate (SAR), ensuring compliance with exposure limits. The study further provides insight into the influence of key parameters on field behavior and the associated trade-offs between focality, penetration, and safety in physiologically relevant stimulation. To the authors' knowledge, this is the first study to combine iTR and TI optimization for microwave-based DBS in a realistic voxel head model, establishing a promising framework for safe non-invasive deep brain stimulation.
\end{abstract}


\section{Introduction}
\label{sec:introduction}
\begin{figure}[b!]
    \centering
    \subfloat[]{\includegraphics[width=0.35\linewidth]{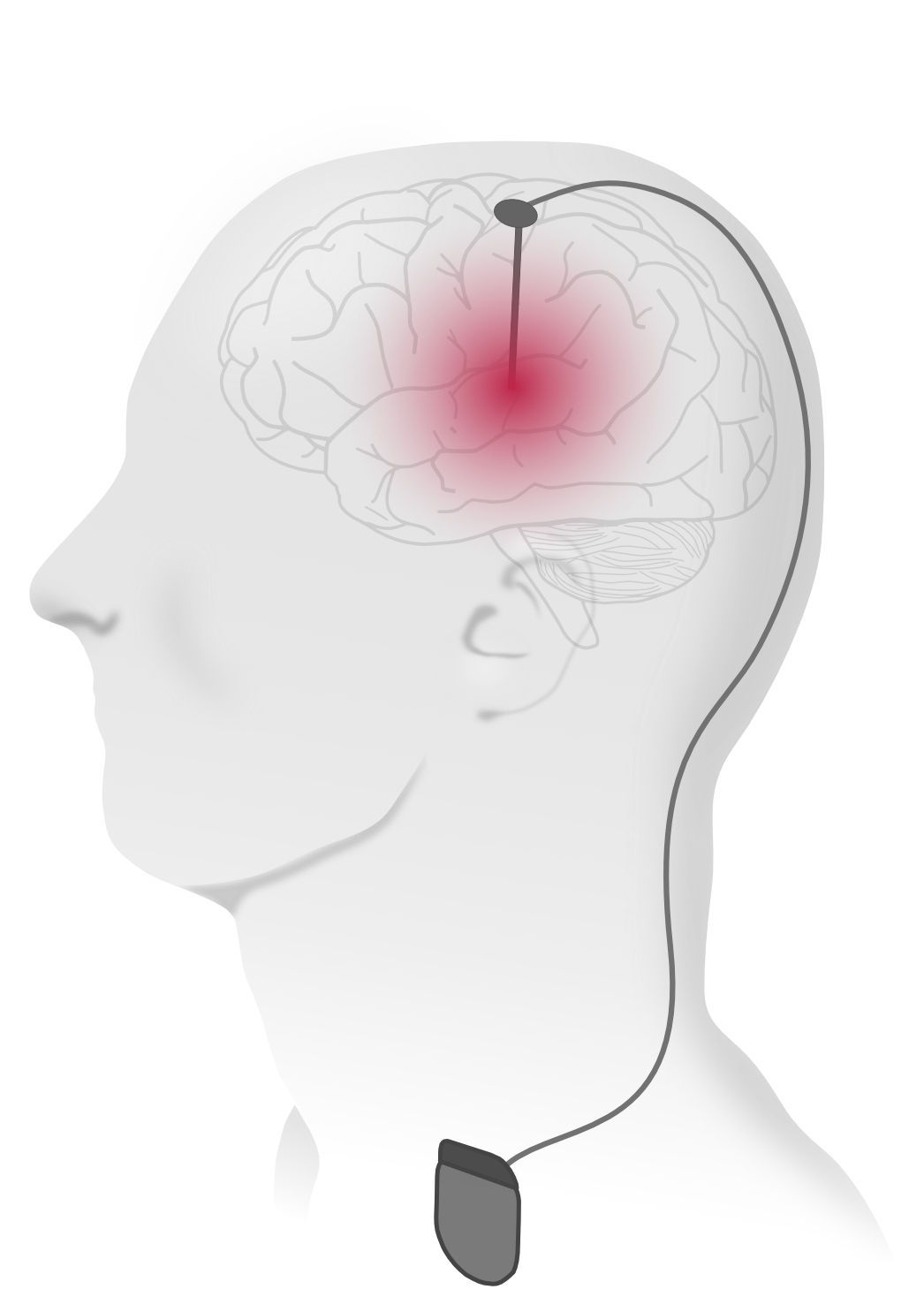}
    \label{fig:DBS_invasive}}
    \hfil
    \subfloat[]{\includegraphics[width=0.35\linewidth]{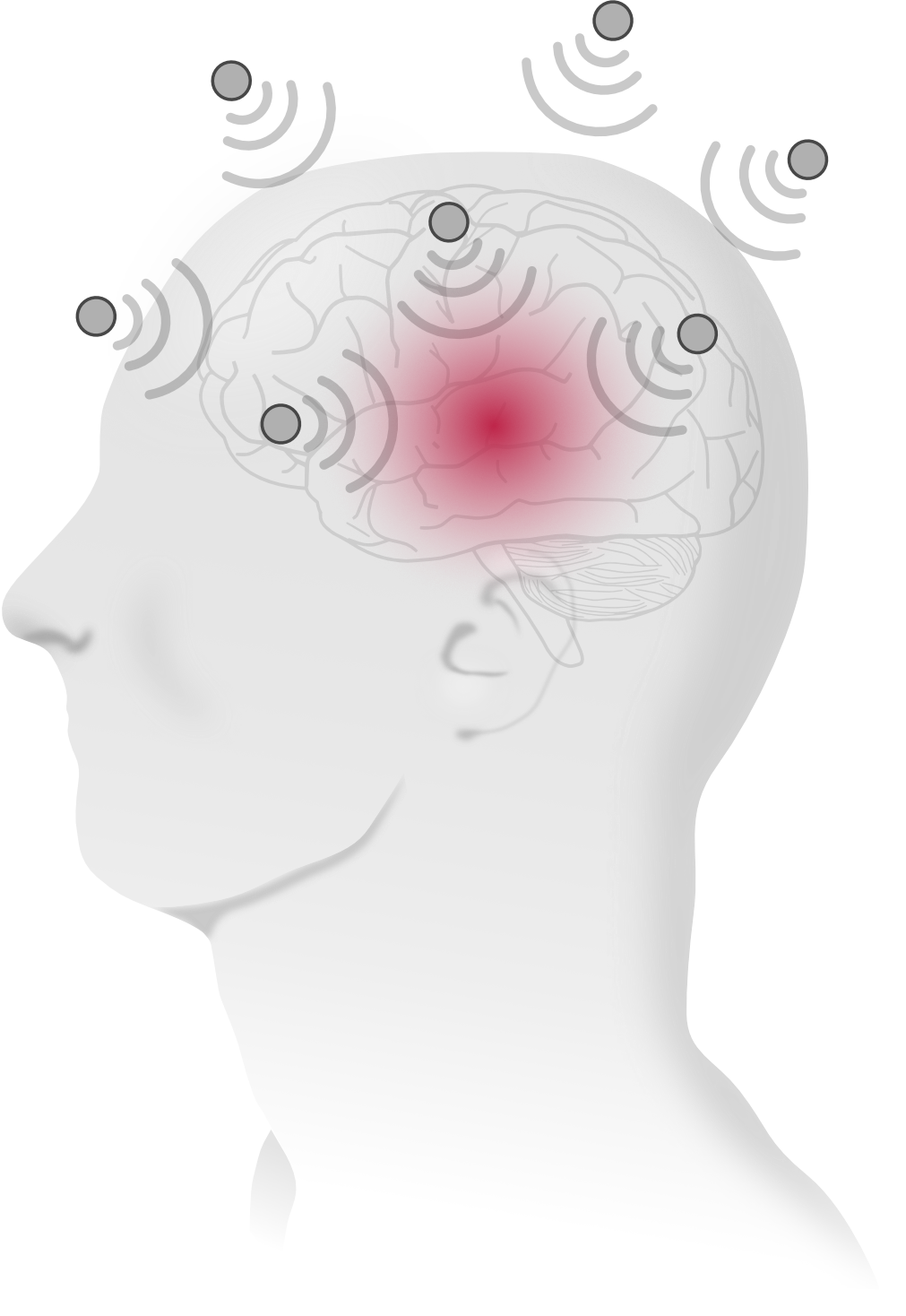}
    \label{fig:DBS_non_invasive}}
    
    \caption{Conceptual illustration of DBS where electric fields induce neural stimulation in a target brain region. (a) The clinically accepted invasive treatment, with implanted electrodes and a pulse generator that generate the field. (b) The proposed non-invasive alternative, where an external MW antenna array generates the field.}
    \label{fig:DBS}
\end{figure}

\IEEEPARstart{D}{eep} Brain Stimulation (DBS) is an established treatment for neurological disorders such as Parkinson’s disease, epilepsy, and depression \cite{Lozano2019, Salanova2018, Wu2021}. Conventional DBS uses implanted electrodes to deliver electrical stimulation to deep brain structures \cite{Krauss2020}, see Fig. \ref{fig:DBS_invasive}. In general, the primary targets for DBS are deep brain structures \cite{Harmsen2020}. Although clinically effective, this invasive approach carries risks including hemorrhage, infection, tissue reactions, hardware failure, and MRI-related complications \cite{Krauss2020, Cheyuo2024}. These limitations motivate the development of non-invasive alternatives.

Existing non-invasive neuromodulation methods such as transcranial magnetic stimulation (TMS) and transcranial electric stimulation (TES) use low-frequency fields that can penetrate the skull, but they provide limited spatial selectivity for deep brain targets \cite{Koponen2020, Truong2020}. Microwave (MW) fields have therefore been proposed due to their potential for improved spatial focusing \cite{Harid2023}. MW systems have already been investigated for medical imaging and therapy applications, including stroke detection, tumor imaging, monitoring of neurological diseases, and hyperthermia treatment \cite{Origlia2024, Guo2023, Ozmen2026, Akazzim2024, Zanoli2021, Rodrigues2021, Zanoli2023}. Together with recent advances in MW brain stimulation \cite{Pereira2024}, these developments suggest that external MW antenna arrays, see Fig. \ref{fig:DBS_non_invasive}, may provide a promising route toward focused non-invasive DBS.

In general, electrical neuromodulation is not achievable at high frequencies due to the neuron’s ability to low-pass filter electrical signals \cite{Hutcheon2000}, with cut-off frequencies typically below \SI{1}{\kilo\hertz} \cite{Buzski2012}, far lower than the radio- or MW frequency range. However, Temporal Interference (TI) overcomes this limitation by combining two high-frequency signals, thereby generating an amplitude-modulated (AM) low-frequency envelope that is able to stimulate neurons \cite{Grossman2017}. By steering the maxima of this envelope, stimulation can be targeted to specific regions inside the head. Since neurons do not respond to the high-frequency components, the fields can pass through outer brain regions without inducing unintended stimulation \cite{Gunduz2020}. Consequently, combining MW focusing and TI techniques for an external MW antenna array could enable targeted neural stimulation, while reducing both off-target stimulation and undesired tissue heating. 

General design principles for phased arrays and field focusing in medical applications have been established in the literature \cite{Bucci2016}. The setup for MW-DBS will be similar to that used in stroke detection and MW hyperthermia treatments, with an antenna array positioned in a matching medium to reduce reflections and improve field transmission into the head \cite{Scapaticci2018, Seebass2001}. Achieving precise field focusing inside the head requires solving complex EM wave propagation problems involving heterogeneous, lossy, and dispersive biological media, which can be effectively addressed using optimization algorithms. EM time reversal (TR) can provide physically informed initial parameters for the optimization of antenna placements and excitations \cite{Madannejad2020, Hajiahmadi2022, deRosny2008}. Due to the brain’s strongly heterogeneous and lossy structure, a single TR solution may yield poorly confined or asymmetric focusing, motivating the use of iterative time-reversal (iTR) to iteratively reduce undesired interference \cite{Zanoli2021}. After initialization, the iTR-derived parameters can be refined using heuristic approaches, such as Particle Swarm Optimization (PSO) \cite{Elkayal2021, Zanoli2023} and Genetic Algorithms (GA) \cite{Baskaran2023}, to determine the optimal antenna excitations, including orientation, amplitude, and phase.

Existing research relevant to this work can be broadly divided into four categories. (i) Microwave hyperthermia, (ii) iTR focusing methods, (iii) MW-based non-invasive DBS focusing, and (iv) TI stimulation studies. Hyperthermia and iTR-based methods aim to maximize absorbed power or field intensity, with formulations inherently tied to thermal or intensity metrics that cannot be directly related to neural stimulation \cite{Zanoli2021, Zanoli2023, Madannejad2020, Hajiahmadi2022, Rodrigues2021}. Similarly, earlier MW-DBS focusing studies optimized single-frequency field magnitudes, often using simplified head models, without incorporating a physiologically relevant stimulation mechanism \cite{Safar2023, Harid2023, LeeW2024, Madannejad2020, Valente2009}. In contrast, TI studies provide a framework for non-invasive neural stimulation, but are largely restricted to low-frequency (kHz) carrier signals, limiting spatial resolution and not accounting for microwave field focusing \cite{Grossman2017, Botzanowski2025}. One study has combined MW focusing with TI optimization \cite{Ahsan2022}. However, it relies on implanted antennas within the skull and therefore does not represent a truly non-invasive approach.

The present work bridges these research directions by introducing a unified iTR-TI framework for non-invasive DBS using MW-TI fields, representing the first computational demonstration of such an approach. The proposed method uses a two-stage framework combining iTR and GA-based TI optimization. Here, iTR determines antenna placement and orientation while providing a physically informed initialization, and TI optimization adjusts frequency assignment, amplitudes, and phases to maximize modulation depth for neural stimulation. In contrast to prior MW focusing studies for DBS that optimize field intensity, the present work introduces a TI-based objective function that explicitly accounts for the neuromodulation mechanism. Furthermore, the framework extends TI to the MHz–GHz regime, enabling compatibility with MW focusing while departing fundamentally from prior kHz-based TI approaches. The proposed method employs an external antenna array of magnetic point dipoles (MPDs) in combination with a realistic MRI-derived voxel head model comprising 38 tissue types, incorporating anatomical heterogeneity and attenuation effects. Its performance is evaluated through a comprehensive numerical validation, including parameter studies, perturbation analysis, and statistical evaluation across multiple configurations, demonstrating a robust and localized AM field with high intensity while adhering to safety constraints. Although direct quantitative comparison with prior work is limited due to fundamental differences in objective functions and reported metrics across prior MW focusing, hyperthermia, and TI studies, the obtained AM fields in this work are higher than those reported in the invasive TI-based MW study \cite{Ahsan2022}. To the best of the authors' knowledge, the proposed framework represents the first integration of iTR and TI optimization for non-invasive MW-DBS in a realistic head model and therefore constitutes an foundation for future development of non-invasive microwave neuromodulation technologies.

\section{Methodology}
We employ a two-stage methodology. First, iTR is used to optimize antenna placement, determining the positions, orientations, and initial steering parameters of the antenna elements. Second, TI optimization refines amplitude, phase, and frequency assignment of each element to maximize the envelope amplitude within the focal region. All field computations are carried out using COMSOL Multiphysics\textsuperscript{\textregistered} \cite{COMSOL_v63}, with optimization performed in MATLAB\textsuperscript{\textregistered} \cite{MATLAB}.

\subsection{Iterative Time Reversal}
\subsubsection{Classical Time Reversal}
The TR operator $\Gamma$ reverses the temporal evolution of electromagnetic fields. In the time domain, it can be mathematically expressed as
\begin{equation}
  \Gamma\{\E(\rvec,t)\} =  \E(\rvec,-t) ,\,\, 
  \Gamma\{\Hvec(\rvec,t)\} = -\Hvec(\rvec,-t), 
\end{equation}
and in the frequency domain, using the time convention $\e^{\imj \omega t}$, 
\begin{equation}
  \Gamma\{\E(\rvec,\omega)\} =  \E^*(\rvec,\omega),\,\,
  \Gamma\{\Hvec(\rvec,\omega)\} = -\Hvec^*(\rvec,\omega). 
\end{equation}
Within an enclosed surface, the EM field can in the lossless case be reconstructed anywhere inside the volume by placing infinitesimal sources on the enclosing surface \cite{derosny2010}.

Although ideally reconstructed over a continuous surface, the field is in practice reconstructed only at discrete locations. To reconstruct the magnetic field at a specific point $\rpvec$, a magnetic dipole $\mvec$ can be placed at that location, with its steering parameters defined by the computed $\Hvec$-field as
\begin{equation}
   \mvec(\rpvec) = -\eta \Hvec^{*}(\rpvec)
\end{equation}
where $\eta$ is the wave impedance \cite{deRosny2008}.

\subsubsection{Forward Field}
\begin{figure}[t]
    \centering
    \includegraphics[width=0.35\linewidth]{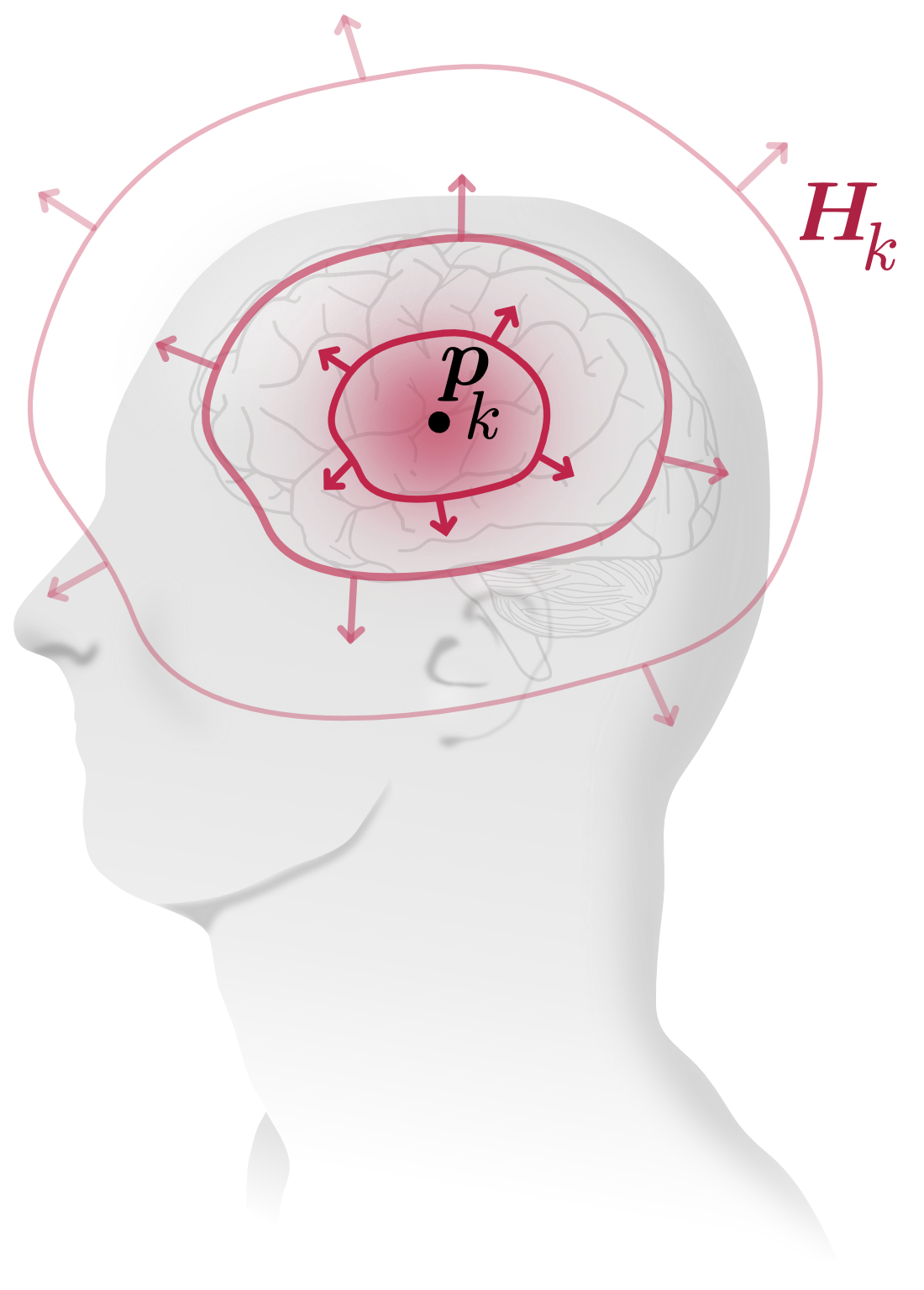}
    \caption{Conceptual illustration of the TR-based field focusing method, numerically implemented in COMSOL. A point source $\pvec_k$ is placed inside the head, generating a forward magnetic field $\Hvec_k$. This is done for $\pvec_k$, $k\in\{x,y,z\}$, giving the total forward field $\Hvec=\sum_k^{x,y,z} \Hvec_k$.}
    \label{fig:method_iTR}
\end{figure}

In this work, a point source is first placed at the desired focal point $\rfvec$ inside the head, and generates the forward field within the head that propagates outward toward the surrounding region. The forward field is computed using the full-wave solver COMSOL. Since COMSOL does not provide an ideal point source, the excitation is approximated using three orthogonal electric point dipoles $ \pvec_k=p_k\phat_k \e^{\imj \alpha_k}$ with $\phat_k  \in \{\xhat,\,\yhat,\,\zhat\}$, where $p_k$ is the amplitude and $\alpha_k$ is the phase. The field from each dipole is computed separately, and the total forward field is obtained by superposition, $\Hvec=\sum_k^{x,y,z} \Hvec_k$, as illustrated conceptually in Fig.~\ref{fig:method_iTR}.

\subsubsection{Backward Field}
The backward field is defined as the complex conjugate of the forward field and evaluated at observation points to determine the external antenna elements. Each element is modeled as an MPD, characterized by its position $\rvec_n$, magnitude $m_n$, dipole moment orientation $\mhat_n$ and phase $\alpha_n$ according to
\begin{equation}
    \mvec_n = m_n \mhat_n \, \e^{\imj \alpha_n}.
\end{equation}
The index $n$ is attributed to the MPDs of the external array, not to be confused with the index $k$ attributed to the point source generating the forward field. Here, the orientation $\mhat_n$ is defined by a unit vector and parameterized using spherical angles $(\theta,\varphi)$, allowing continuous orientation in three dimensions. Throughout this manuscript, any reference to antenna orientation specifically denotes the unit vector of the magnetic dipole moment $\mhat_n$, rather than a geometrical or radiation-pattern direction.

The TR algorithm calculates the local weighted energy at all candidate antenna positions $\rvec_i$ according to
\begin{equation}
    W(\rvec_i) = \sum_o w(\rvec_o, \rvec_i)\,
    \abs{\beta(\rvec_i, \rfvec)\,
    \Hvec^*(\rvec_i)}^2
\end{equation}
where $\rvec_o$ are observation points around each $\rvec_i$, and $w(\rvec_o, \rvec_i)$ a sampling weight used to evaluate the local field around each candidate dipole. It is defined as a Gaussian function decaying with distance from the candidate antenna position according to
\begin{equation}
    w(\rvec_o,\rvec_i) = \exp{-\frac{\abs{\rvec_o-\rvec_i}^2}{2\sigma^2}}
\end{equation}
where $\sigma$ is the standard deviation of the Gaussian weighting function. 

Then, the candidate positions are ranked based on the highest energy. Each MPD is assigned the position $r_n$ of the candidate with the highest energy, after which the algorithm minimizes the following TR objective function 
\begin{equation}
    \mathcal{J}_\mathrm{TR} = \sum_o \sum_n w(\rvec_o, \rvec_n)\left|
    \mvec_n
    - \beta( \rvec_n, \rfvec)^b\,
      \Hvec^*(\rvec_n)\,
    \right|^2
    \label{eq:obj_TR}
\end{equation}
where $\mvec_n$ are the selected dipoles at positions $\rvec_n$. The factor $\beta(\rvec_n, \rfvec)$ is an attenuation compensation factor and represents the inverse of the expected attenuation along a straight path between the MPD position $\rvec_n$ and focal point $\rfvec$. It is defined as
\begin{equation}
    \beta(\rvec_n, \rfvec) = \exp{\int_{\rvec_n}^{\rvec_\mathrm{f}}\alpha(\rvec) \, \diff \ell}
\label{eq:beta_func}
\end{equation}
where the local attenuation constant $\alpha(\rvec)$ is given by
\begin{equation}
    \alpha(\rvec) = \omega \sqrt{\frac{\mu_0 \varepsilon_0}{2}}  \sqrt{|\varepsilon(\rvec)| - \mathfrak{R}\{\varepsilon(\rvec)\} }
\end{equation}
which represents a high-frequency approximation of the local attenuation in a lossy dielectric medium. The exponent $b\in[0,1]$ in Eq. (\ref{eq:obj_TR}) is introduced to control the degree of attenuation compensation applied through $\beta^b$. Selecting $b$ within this range therefore allows a balanced, partial correction for tissue losses while avoiding large weighting of distant or heavily attenuated dipoles. 

\subsubsection{Iterative Time Reversal}
When performing TR iteratively, the point source steering parameters and the attenuation-compensation exponent $b$ are updated. In each iteration, a forward field is computed, a backward field is reconstructed, and the parameter set $\zetavec = [\{p_k\}_k,\,\{\alpha_k\}_k,\,b \big]$ is optimized to obtain a backward field that best approximates the ideal TR field. The iTR objective function is formulated in terms of the electric field since neural stimulation is determined by the local electric field rather than the magnetic field. To formalize this optimization, we define the iTR objective function for the backward field as
\begin{equation}
    \mathcal{J}_\mathrm{iTR} = c_1 \text{HIR} + c_2 \text{FIQ}
    \label{eq:obj_iTR}
\end{equation}
with positive weights $c_1$ and $c_2$. The half-intensity radius (HIR) is the smallest radius for which the spherical average of the intensity of the electric field, $I(\rvec) = \norm{\E(\rvec)}^2$, falls below half of its value at the focal point. Here, $||\cdot||$ denotes the Euclidean norm of the field. The focus intensity quotient (FIQ) evaluates field confinement and is defined as
\begin{equation}
    \text{FIQ} = \frac{\overline{I(\rvec)}_{\Omega_\mathrm{b}}}{\overline{I(\rvec)}_{\Omega_\mathrm{f}}}
    \label{eq:FIQ}
\end{equation}
where $\Omega_\mathrm{f}$ is the focal region and $\Omega_\mathrm{b}$ is the background region. The objective function Eq. (\ref{eq:obj_iTR}) therefore promotes a narrow focal region while minimizing unintended field levels elsewhere.

\begin{figure}[!t]
  \centering
  \includegraphics[width=\linewidth]{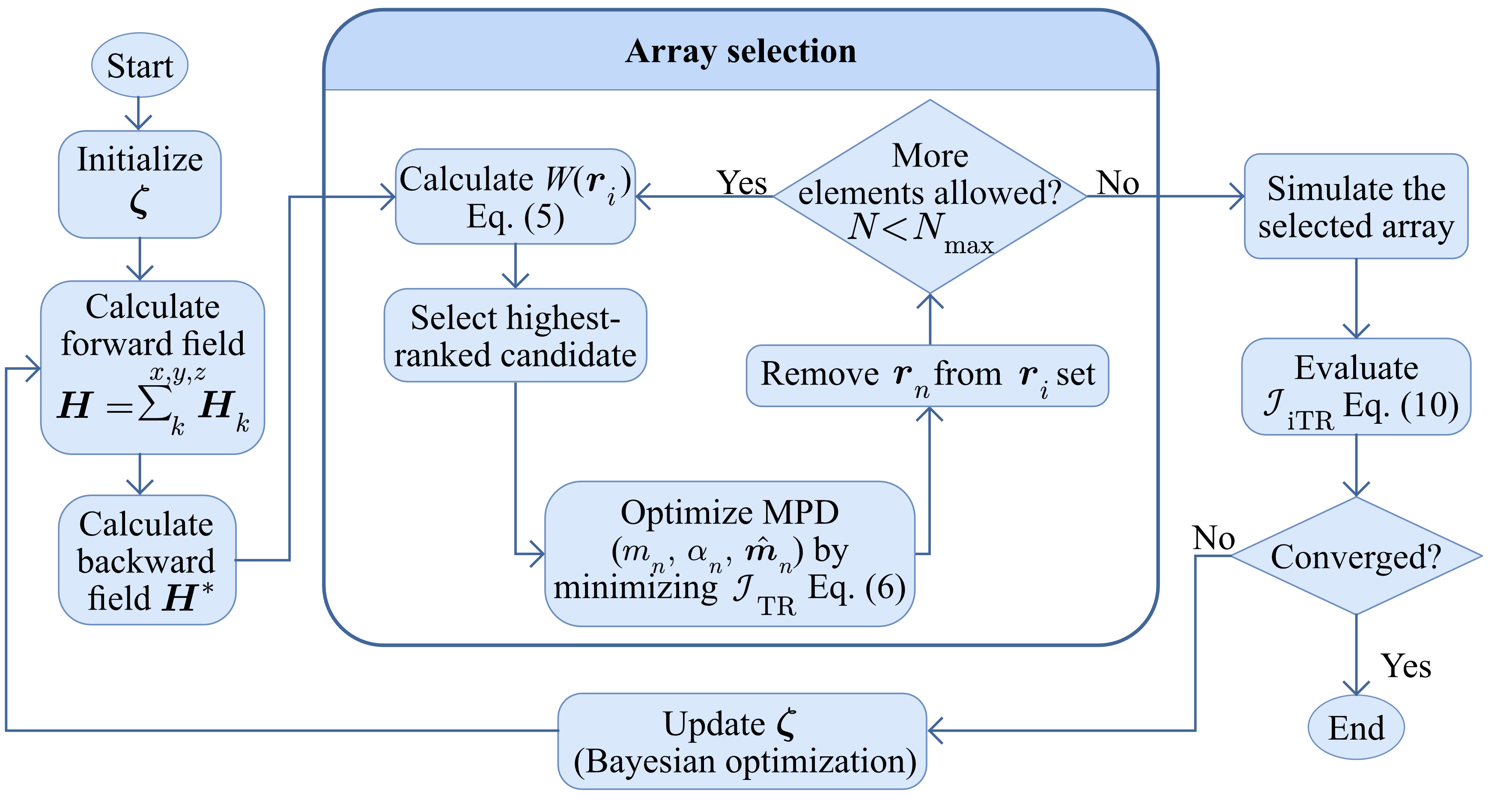}
  \caption{Flowchart showing the implemented iTR optimization.}
  \label{fig:ITR_flowchart}
\end{figure}

The iTR algorithm, illustrated in Fig. \ref{fig:ITR_flowchart}, iteratively constructs the external antenna array by alternating between forward-field simulation, backward-field reconstruction, optimization of $\zetavec$, with updates guided by the iTR objective function Eq. (\ref{eq:obj_iTR}) until convergence. 

\subsection{Temporal Interference}
\label{sec:TI}
\begin{figure}[!t]
    \centering
    \includegraphics[width=0.75\linewidth]{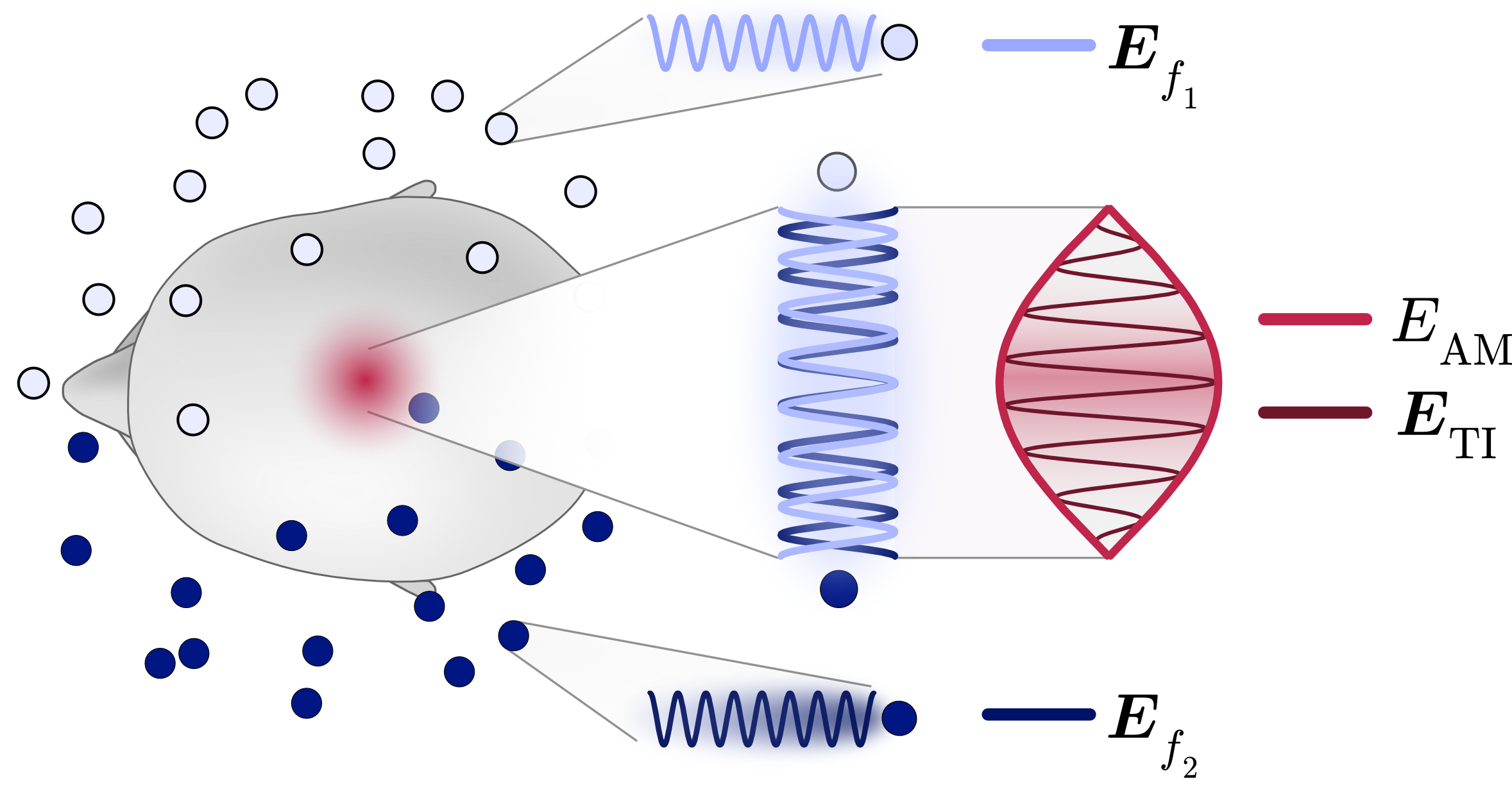}
    \caption{Conceptual illustration of TI stimulation. The antenna array (dots) generates two high-frequency fields $\E_{f_1}$ and $\E_{f_2}$ (light and dark blue), that interfere at a deep brain target region to produce an AM field $\E_\mathrm{TI}=\E_{f_1}+\E_{f_2}$ with a low-frequency envelope $\EAM$ (red).}
    \label{fig:method_TI}
\end{figure}

The neural stimulation mechanism underlying TI was first introduced in \cite{Grossman2017}, where it was hypothesized that neurons demodulate AM signals, responding to the low-frequency envelope rather than the high-frequency carrier. The AM field is obtained by combining two electric fields $\E_\mathrm{TI} = \E_{f_1} + \E_{f_2}$ where $\E_{f_1}$ and $\E_{f_2}$ denote the fields generated at frequencies $f_1$ and $f_2=f_1+\Delta f$, respectively. Their superposition produces an AM field whose envelope oscillates at the difference frequency $\Delta f$, as illustrated in Fig. \ref{fig:method_TI}.

Under this TI hypothesis, neural activation is expected to occur in regions where the modulation depth of the electric field is maximal. The envelope amplitude at a point $\rvec$, as defined in \cite{Ahsan2022}, is given by
\begin{equation}
    E_{\mathrm{AM}}(\rvec) = 2\,\min\big( \norm{\E_{f_1}(\rvec)}\,,\, \norm{\E_{f_2}(\rvec)} \big).
    \label{eq:EAM}
\end{equation}
This low-frequency envelope is believed to drive neural modulation even when the carrier frequencies lie in the \si{\mega\hertz} or \si{\giga\hertz} frequency range \cite{Beason2002, Ahsan2023}. Thus, maximizing the envelope amplitude, defined in Eq. \eqref{eq:EAM}, within the focal region is a central objective in TI field optimization.

After obtaining an optimized array geometry from the iTR algorithm, the antenna positions $\rvec_n$ and orientations $\mhat_n$ are fixed. The magnitudes $m_n$ and phases $\alpha_n$ are then used as initial values for the TI optimization. Each antenna element is assigned one of the two operating frequencies $f_1$ or $f_2$, and its amplitude and phase at the assigned frequency are independently adjustable, denoted as $m_{n_i}$ and $\alpha_{n_i}$, where $i\in\{1,2\}$ indicates the frequency index. The resulting electric field contribution from element $n_i$ at frequency $f_i$ is described by its Green’s function $\mathbf{G}_{n_i}$. The total electric field at frequency $f_i$ is
\begin{equation}
    \E_{f_i} = \sum_{n_i=1}^{N_i} \mathbf{G}_{n_i}(f_i)m_{n_i} \e^{\imj\alpha_{n_i}}.
    \label{eq:total field 2 freq}
\end{equation}
Here, $N_i$ is the number of elements assigned to frequency $f_i$. Each $\mathbf{G}_{n_i} \in \mathbb{C}^3$ contains the three Cartesian field components and is linear with respect to amplitude and phase, making it suitable for frequency-domain optimization \cite{Harid2023, Ahsan2022}.

\begin{figure*}[!t]
  \centering
  \includegraphics[width=0.95\textwidth]{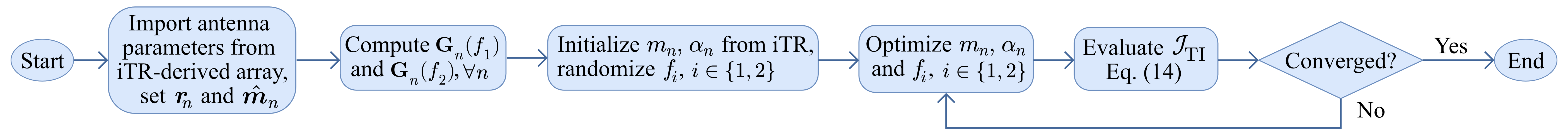}
  \caption{Flowchart showing the implemented TI optimization. }
  \label{fig:TI_flowchart}
\end{figure*}

To localize the TI envelope at the target region, the optimization minimizes the following objective function
\begin{equation}
    \mathcal{J}_\mathrm{TI} = \frac{\overline{E_\mathrm{AM}(\rvec)}_{\Omega_\mathrm{b}}}{\overline{E_\mathrm{AM}(\rvec)}_{\Omega_\mathrm{f}}} 
    \label{eq:obj_TI}
\end{equation}
which measures the ratio of the average TI modulation depth in the background region $\Omega_\mathrm{b}$ to that in the focal region $\Omega_\mathrm{f}$. Minimizing $\mathcal{J}_\mathrm{TI}$ concentrates the spatial localization of the TI envelope in the target while suppressing it elsewhere. Since high modulation depth requires both large and balanced field amplitudes from $f_1$ and $f_2$, the objective is sensitive to both relative and absolute field strengths. The TI algorithm is illustrated in Fig. \ref{fig:TI_flowchart}.

\subsection{Field Evaluation Metrics}
\label{sec:focality_intensity}
The HIR and FIQ metrics provide a measure of single-frequency field focusing in the iTR stage. However, they cannot be applied in the TI stage, since the relevant quantity for TI arises from the interference of two frequency fields, rather than the magnitude of a single-frequency field. To quantify the spatial confinement and field strength of the AM field, we therefore define separate metrics for focality and intensity. 

The focality volume, denoted $\Vfoc$, is defined as the connected set of voxels satisfying
\begin{equation}
    \EAM(\rvec) \geq 0.5\,\EAMmax,
\end{equation}
within the neural tissue domain $\Omega_{\mathrm{n}}$, and containing the target focus location $\rfvec$. Here,
\begin{equation}
    \EAMmax = \max \EAM(\rvec \in \Vfoc)
\end{equation}
denotes the maximum AM field magnitude within the focality volume, referred to as the focal intensity. We now introduce a physical volume of $\Vfoc$ denoted $V_{50\%}$ (m$^3$). The focality, $F$, is defined as
\begin{equation}
    F = V_{50\%}^{1/3}.
\end{equation}
Thus, the focality $F$ quantifies spatial confinement, with lower values corresponding to better confined fields and higher values indicating more spatially extended fields.

In addition to the focal intensity, we quantify the field strength in the surrounding neural tissue by defining the background intensity. This is evaluated as the maximum of $\EAM$ in the region $\Omega_\mathrm{n} \setminus \Vfoc$ i.e., the neural tissue outside the focality volume. For brevity, $\rvec_F$ and $\rvec_\mathrm{B}$ denote positions in the focality volume and the background region, respectively. This distinction allows the intensity to be assessed both within the focality volume, and in the background neural tissue, where modulation is undesired.

Since $\Vfoc$ depends implicitly on the unknown quantity $\EAMmax$, a self-consistent solution is obtained using a two-step procedure. First, an initial estimate of the intensity is taken as the field value at the focus location, i.e., $E_0 = \EAM(\rfvec)$. This value is used to define a threshold, from which a connected volume containing $\rfvec$ is identified. The maximum field value within this volume is then extracted and used to update the intensity estimate. In a second step, the threshold is recomputed using this updated value, and the focality volume is re-evaluated. This yields the final values of $\Vfoc$, $\EAMmax$, and $F$ and ensures that they are defined consistently with respect to the true peak field value.

\subsection{Specific Absorption Rate}
\label{sec:method SAR}
Human exposure to EM fields is regulated by international safety standards, most notably the IEEE guidelines \cite{IEEE_safety} and ICNIRP \cite{ICNIRP2020}. In this work, safety compliance is evaluated in terms of SAR in accordance with the IEEE guidelines for EM exposure assessment \cite{IEEE_safety}. The local SAR is defined as
\begin{equation}
    \mathrm{SAR}(\rvec) = \frac{\sigma(\rvec) ||\E(\rvec)||^2}{2 \rho(\rvec)}.
    \label{eq:SAR}
\end{equation}
Here, $\E$ denotes the peak electric field magnitude obtained from the frequency-domain solution, whereas the spatially varying electrical conductivity $\sigma$ (S/m) and mass density $\rho$ ($\mathrm{kg/m}^3$) are taken from the IT’IS database \cite{ITISTissueDatabase}.

For TI excitation, the SAR contribution from each driving frequency is computed independently and subsequently summed to obtain the total SAR exposure, consistent with IEEE multi-frequency exposure standards \cite{IEEE_safety}. Compliance is evaluated based on the 10-g spatially averaged SAR, denoted $\SAR$, calculated over a cubic region, with an upper limit of 10~W/kg.

\section{Simulation Environment}
A human head model was placed inside a spherical domain with a scattering boundary condition on its outer surface. MPDs were positioned around the head, and a matching layer was added between the head model and the scattering boundary to minimize reflections from the radiating antennas.

\begin{figure}[b!]
  \centering
  \includegraphics[width=0.5\textwidth]{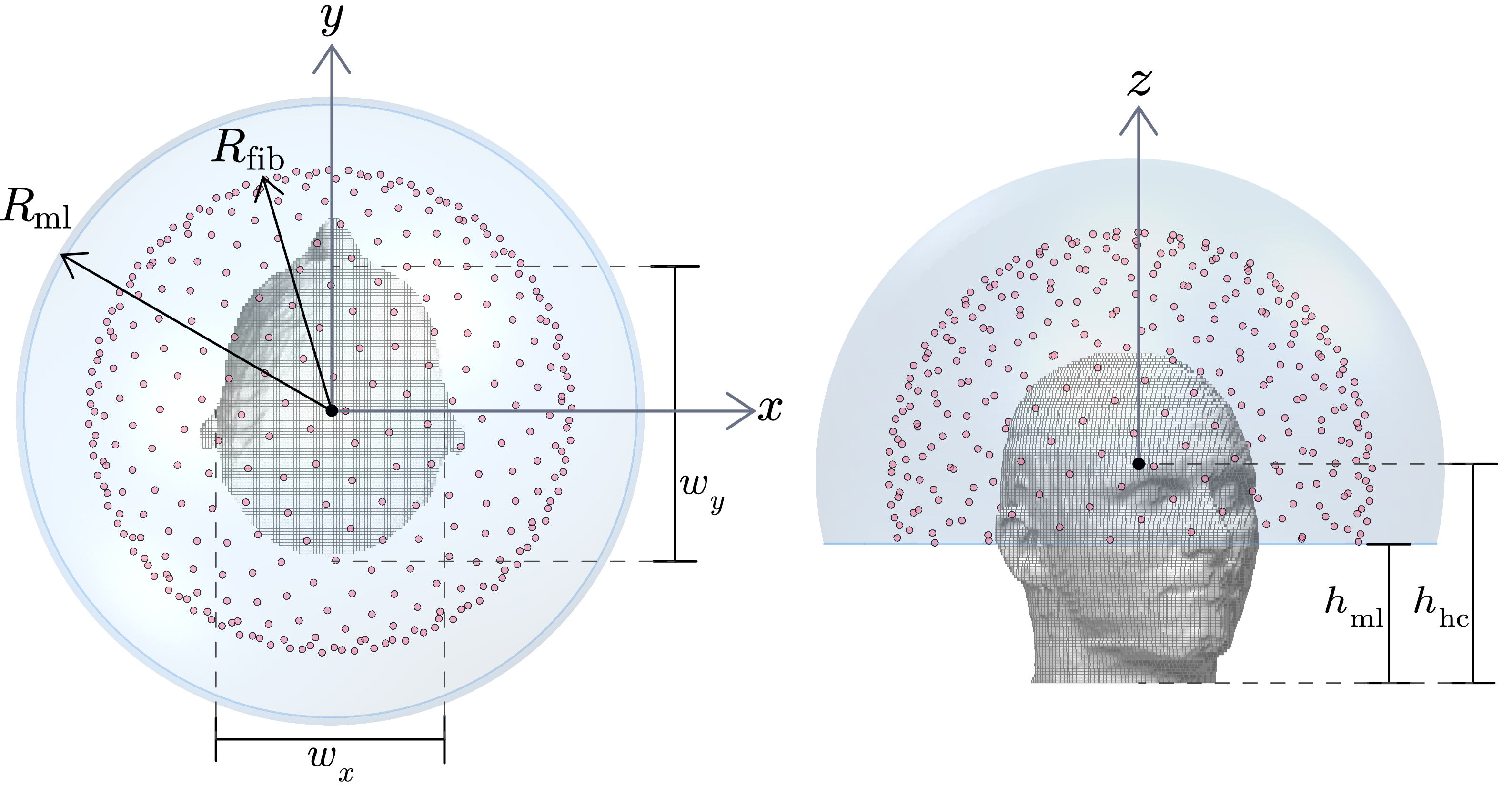}
  \caption{Geometric configuration of the voxelized Duke head model and antenna search space. The head model is enclosed by a matching layer (light blue), with candidate antenna positions $\rvec_i$ (pink dots) distributed on the Fibonacci sphere prior to iTR optimization. The head center at (0,0,0) is indicated by a black dot. The head width is denoted by $w_x$ and $w_y$, the Fibonacci sphere radius by $R_\mathrm{fib}$, and the matching layer radius by $R_\mathrm{ml}$. The vertical distances from the neck crop to the head center and to the matching layer cutoff are given by $h_\mathrm{hc}$ and $h_\mathrm{ml}$, respectively.}
  \label{fig:Sim_setup}
\end{figure}

An illustrative overview of the simulation configuration is shown in Fig.~\ref{fig:Sim_setup}. The figure shows the voxelized head model centered within the matching layer and the surrounding spherical antenna placement domain. Candidate antenna positions are distributed on an enclosing surface, and the coordinate system and key geometric dimensions of the model are indicated. The individual parameters and their definitions are introduced in the following sections.

\subsection{Anatomical Head Model}
The human head is highly inhomogeneous and anisotropic, consisting of numerous anatomical structures and tissue types with a great variation in dielectric properties. This complexity makes accurate EM modeling particularly challenging. Voxelized anatomical models address this challenge by providing detailed, spatially resolved representations of biological tissues. Among these, the IT’IS Foundation models are widely used and well validated in bioelectromagnetic research. In the present work, we used the anatomical head model Duke, from the IT'IS Virtual Population \cite{gosselin2014virtual}. 

The head was isolated from the full-body model by truncation at a height of 1.6~m. The head center was then defined as the origin of the coordinate system, $\rvec_\mathrm{hc} = (0,0,0)$, which corresponds to a position $h_\mathrm{hc}=\SI{15}{\centi\meter}$ above the truncation plane. This point is located centrally within the head in the lateral directions, i.e., at the midpoint of the head width dimensions $w_x$ and $w_y$. The resulting spatial configuration, including the coordinate system and characteristic dimensions, is illustrated in Fig.~\ref{fig:Sim_setup}.

The voxel count increases rapidly with finer discretization. To balance computational cost and spatial resolution, a voxel size of $2 \times 2 \times 2$ mm was selected. This resolution is sufficient to capture the dominant propagation effects in the head. The choice is consistent with established practice in bioelectromagnetic modeling, such as hyperthermia treatment planning, where benchmark studies employ head model resolution of 2.5~mm \cite{Paulides2021}. A voxel-based NASTRAN mesh was generated and imported into COMSOL to define the full three-dimensional head geometry.

\begin{figure}[b!]
    \centering
    \subfloat[]{\includegraphics[scale=0.37]{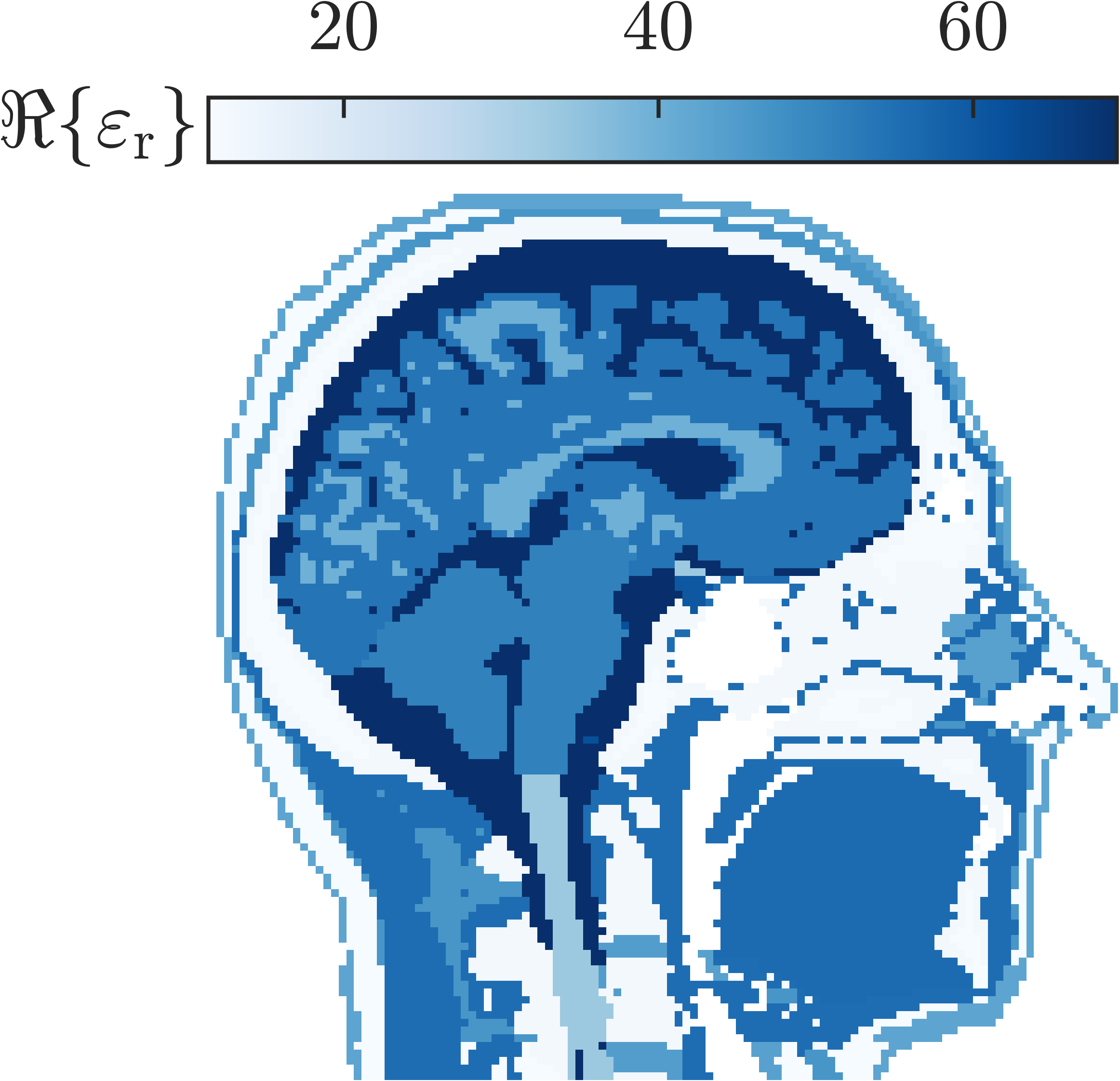}
    \label{fig:permittivity_Re}}
    \hfil
    \subfloat[]{\includegraphics[scale=0.37]{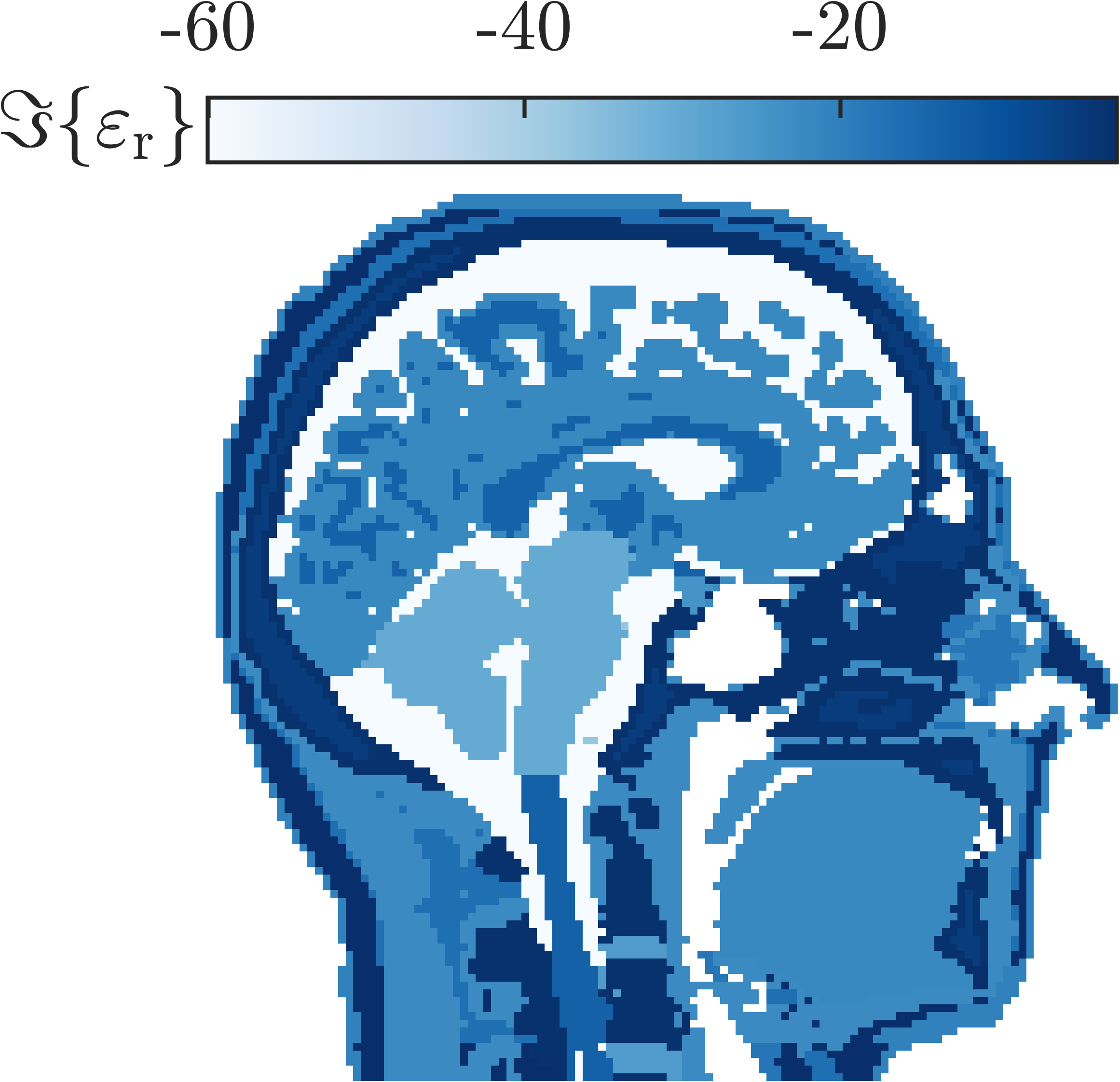}
    \label{fig:permittivity_Im}}
    
    \caption{
    Real (a) and imaginary (b) parts of the relative permittivity $\varepsilon_\mathrm{r}$ in a central slice of the voxelized Duke head model, for a voxel size of \SI{2}{\milli\meter}, evaluated at \SI{700}{\mega\hertz}.}
    \label{fig:permittivity}
\end{figure}

For material characterization, each voxel was assigned an anatomical tissue label from the Virtual Family Duke head model, where each label corresponds to a segmented anatomical structure (e.g., gray matter, white matter, bone, or blood). This resulted in 44 anatomical labels within the head. As some anatomical labels share identical material properties, the final head model consists of 38 unique tissue types. Dielectric properties were assigned using the IT’IS Foundation Tissue Database \cite{ITISTissueDatabase} to model dispersive dielectric behavior, i.e., frequency‑dependent complex permittivity. The materials are assumed to be isotropic, such that their EM response is direction‑independent. Fig.~\ref{fig:permittivity} illustrates the resulting permittivity distribution in a central slice of the head model, with a voxel resolution of \SI{2}{\milli\meter} and evaluated at a frequency of \SI{700}{\mega \hertz}. The figure highlights the strong spatial variation in dielectric properties across tissue types, which will affect the AM field distribution generated by the external array.

\subsection{External Antenna Array}
\label{sec:method antenna array}
MPDs are used to model the antenna elements. The use of idealized sources enables evaluation of the proposed optimization framework independently of antenna geometry, providing a design-independent benchmark. In COMSOL, the MPDs are implemented using the built-in \emph{Magnetic Point Dipole} feature. Each MPD is defined by a complex dipole moment $m_n \e^{\imj \alpha_n}$ and oriented along $\mhat_n$, which together specify the magnitude, phase, and direction.

The model assumes idealized, non-interacting sources and therefore does not account for mutual coupling, realistic radiation patterns or impedance matching. A minimum inter-element spacing $d_\mathrm{min} = \SI{6}{\centi\meter}$ is imposed to limit near-field interactions and avoid unrealistically dense configurations with clustering of elements, serving as a simplified approximation rather than a full treatment of electromagnetic interactions. 

A lossless matching layer with permittivity $\varepsilon_\mathrm{ml} = 40$ was introduced. The matching layer is a spherical volume of radius $R_\mathrm{ml}=\SI{20}{\centi\meter}$, truncated above at $h_\mathrm{ml}=\SI{10}{\centi\meter}$ from where the head is separated from the body, leaving air below to reflect a realistic scenario where the matching layer cannot cover the mouth or nose, see Fig. \ref{fig:Sim_setup}. 

Candidate antenna positions $\rvec_i$ were defined within the matching layer on a Fibonacci sphere with a radius of $R_\mathrm{fib}= \SI{17}{\centi\meter}$, centered at the head center $\rvec_\mathrm{hc}$, providing approximately 300 candidate locations (see Fig. \ref{fig:Sim_setup}). The radius $R_\mathrm{fib}$ corresponds to the distance from $\rvec_\mathrm{hc}$ to the antenna locations, ensuring sufficient spatial coverage while satisfying the imposed minimum inter-element spacing constraint. 

The selection of antenna radius and matching medium parameters is guided by preliminary simulations to ensure adequate field penetration into the head. Detailed modeling of antenna geometry, mutual coupling effects, and matching optimization is beyond the scope of this work, as the focus is on optimizing steering parameters and the resulting envelope field distribution using the proposed iTR–TI framework.

\subsection{Iterative TR Optimization Setup}
The TR stage is configured using a Gaussian weight width $\sigma = \frac{0.8}{3} \, d_\mathrm{min}$, such that the sampling window $w(\rvec)$ is negligible beyond $3\sigma = 0.8d_\mathrm{min}$. This choice ensures minimal overlap between neighboring sampling regions. The TR stage does not require an explicit initial guess, as the antenna array is constructed sequentially based on the time-reversed field, with dipole parameters determined from local field matching.

To further refine the array construction, the iTR procedure updates the forward-field parameters in $\zetavec$, using Bayesian optimization in MATLAB\textsuperscript{\textregistered}. The optimization is initialized with a small set of sample evaluations to explore the parameter space, making it well-suited for this low-dimensional, nonlinear problem. The optimization is performed for a fixed number of seven iterations. In practice, convergence was reached within these iterations for all cases considered, with no further improvement observed.

\subsection{TI Optimization Setup}
PSO-based optimization approaches are common in MW focusing \cite{Zanoli2023,Elkayal2021} but do not natively handle discrete variables. GA is preferred over PSO because it natively supports discrete variables, avoiding the performance-degrading rounding required by PSO's continuous formulation. Therefore, GA is used to enable element-wise assignment to $f_1$ or $f_2$, which is required for TI. This formulation leads to a mixed integer–continuous optimization problem over frequency assignment, amplitudes, and phases.

The optimization variables consist of a concatenated vector containing the frequency assignment (binary), magnitude (non-negative continuous), and phase (continuous and bounded within $[-\pi,\pi]$) for all elements. Antenna positions and orientations are fixed from the iTR solution. The magnitudes and phases are initialized using the iTR-derived steering parameters, while the frequency assignment is initialized randomly to avoid bias and ensure sufficient variation between candidate solutions. This hybrid initialization provides a physically informed starting point while preserving sufficient diversity in the search space, which has been shown to improve convergence compared to purely random initialization.

A population of 300 individuals is used to explore the high-dimensional search space. A fraction of 5\% of the best-performing individuals is preserved between generations to ensure retention of the best solutions. A crossover fraction of 0.8 is used to generate new solutions by recombining high-quality candidates. Mutation introduces controlled variations that adapt during the optimization while ensuring all variables remain within their bounds, maintaining robustness in the mixed integer–continuous setting.

The optimization is terminated when the relative improvement in the best objective value $\JTI$ falls below 1\% over 15 consecutive generations or after a maximum of 300 generations. For all cases considered, the stopping criterion based on relative improvement was satisfied before reaching the maximum iteration limit.

\subsection{Focal Point and Focal Region Definition}
\label{sec:method focal region}
The focal point $\rfvec$, also referred to as target point or location, is defined in a head-centered coordinate system with the center focus located at $\rfvec = \rvec_\mathrm{hc}= (0,0,0)$. Any off-center focus is introduced as a spatial offset from this reference, such that the shifted focal point is expressed as a displacement relative to the head center.

In this work, the focal region $\Omega_\mathrm{f}$ used in Eqs. (\ref{eq:FIQ}) and (\ref{eq:obj_TI}) is defined as a sphere centered at the focal point $\rfvec$. It does not represent a prescribed therapeutic volume or anatomical structure, but rather serves as a numerical support region used in these objective functions to localize the envelope maximum around the focal point. Unlike MW hyperthermia, which requires full tumor coverage, the requirements for TI-based DBS remain incompletely characterized, and the optimal spatial extent of modulation for neural activation is not yet established. Consequently, the optimization framework is formulated to favor maximal spatial confinement of the field rather than volumetric coverage, effectively approximating a point-target objective. The background region $\Omega_\mathrm{b}$ is defined as the rest of the head, excluding the focal region.

With this definition, the achieved focusing performance is assessed in terms of how well the field is confined around the focal point $\rfvec$, as quantified by the focality metric $F$ introduced in section \ref{sec:focality_intensity}.

\section{Results}
We present a broad numerical validation of the proposed framework through systematic parameter sweeps and statistical evaluation across realistic sources of variability. First, a benchmark case is introduced as a reference configuration and used as the baseline for all subsequent comparisons. Using this benchmark, key parameters are systematically varied, including operating frequency, number of antenna elements, focal region definition, and target location within the brain.

For each scenario, we report quantitative performance metrics specifically defined for TI fields, including the focality $F$ (cm), the intensity $\EAMmax$ (V/m) within the focality volume $\Vfoc$, and intensity in the background region $\Omega_\mathrm{n}\setminus \Vfoc$. Additionally, the maximum 10~g averaged specific absorption rate, $\SAR$ (W/kg), is evaluated in relevant tissues, namely skin, cerebrospinal fluid (CSF), and skull. The minimum and maximum values of MPD magnitudes $m$ (\si{\micro \ampere m^2}) required to excite the optimized antenna array are also presented.

Spatial distributions of $\EAM$ and $\SAR$ are also presented. All field distributions are presented in absolute units, enabling direct assessment of field levels and associated safety metrics. Field distributions are visualized in sagittal, coronal, and axial planes intersecting at the focal location $\rfvec$ (marked with a white cross). A diverging colormap is used, with values above and below 50\% of the intensity $\EAMmax$ shown in red and blue, respectively.

The results section concludes with a statistical summary of all simulated cases to assess the overall robustness and variability of the proposed framework. These results allow for assessment of how focusing performance and field levels depend on target definition, anatomical heterogeneity, and array configuration.

\subsection{Benchmark Configuration}
\label{sec:res_base}
To provide a reference for evaluating the proposed framework, a benchmark configuration is defined as follows. The focal region $\Omega_\mathrm{f}$ is defined as a sphere of radius $\SI{1}{\centi\meter}$ centered at the focal point $\rfvec=(0,0,0)$. The operating frequency is set to $f_1 = \SI{700}{\mega\hertz}$ and $f_2 = f_1 + \Delta f$ with $\Delta f = 100$~Hz. This produces a low-frequency modulation envelope at 100~Hz, consistent with values used in other TI studies~\cite{Ahsan2022, Botzanowski2025} and within the range typically used for neural stimulation in DBS~\cite{Su2018}. The array consists of $N=35$ antenna elements. 

To account for variability in the optimized antenna array, multiple simulation runs were performed, each yielding a distinct array realization, with results summarized in Table~\ref{tab:res_base}. Case 6 is selected as the reference for all subsequent parameter sweeps due to its representative performance, being closest to the mean values obtained in the statistical evaluation presented later in the paper, and is used to illustrate the antenna configuration and field distributions.

\begin{figure*}[!t]
    \centering
    \subfloat[]{\includegraphics[width=0.18\textwidth]{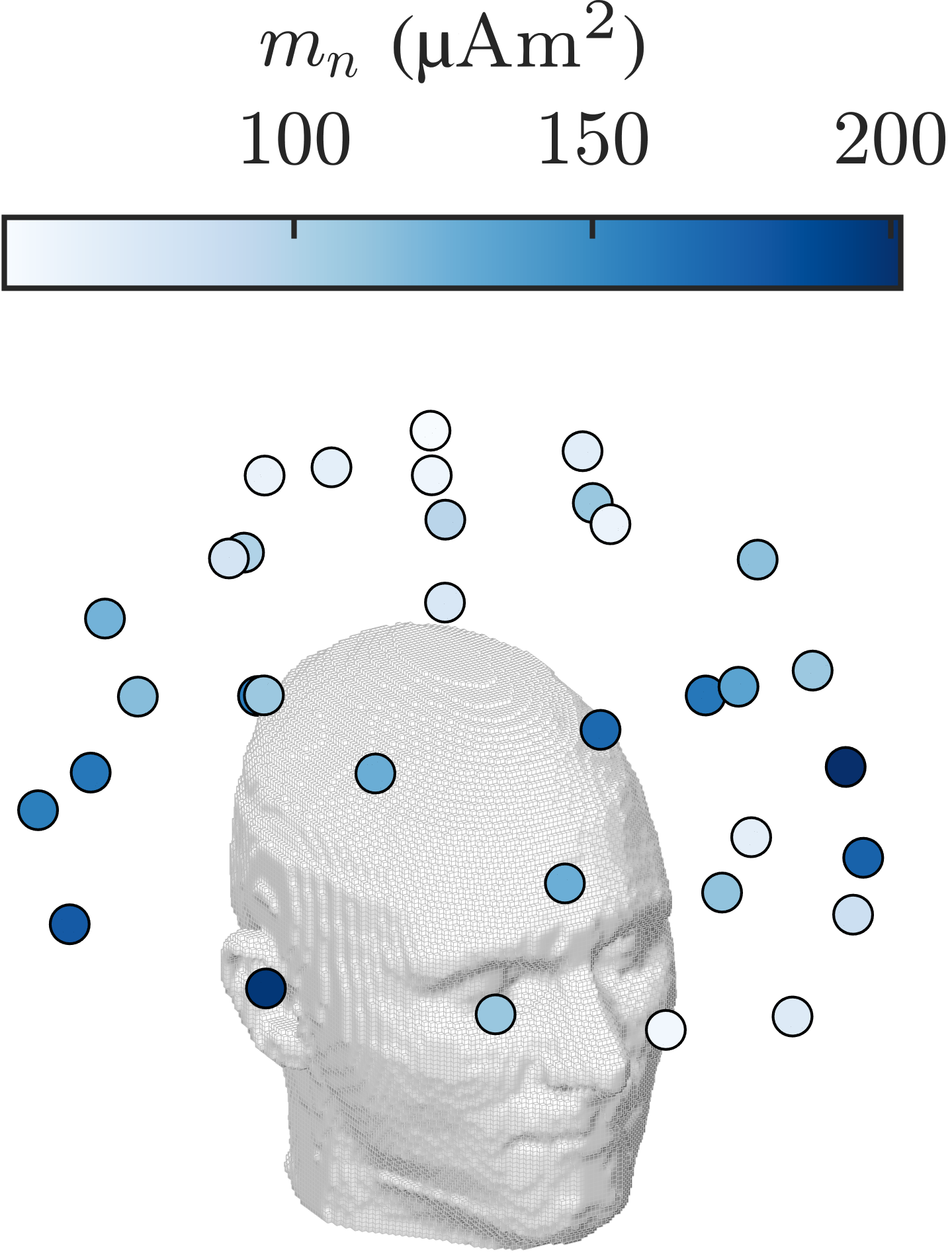}
    \label{fig:res_TI_array_m}}
    \hfil
    \subfloat[]{\includegraphics[width=0.18\textwidth]{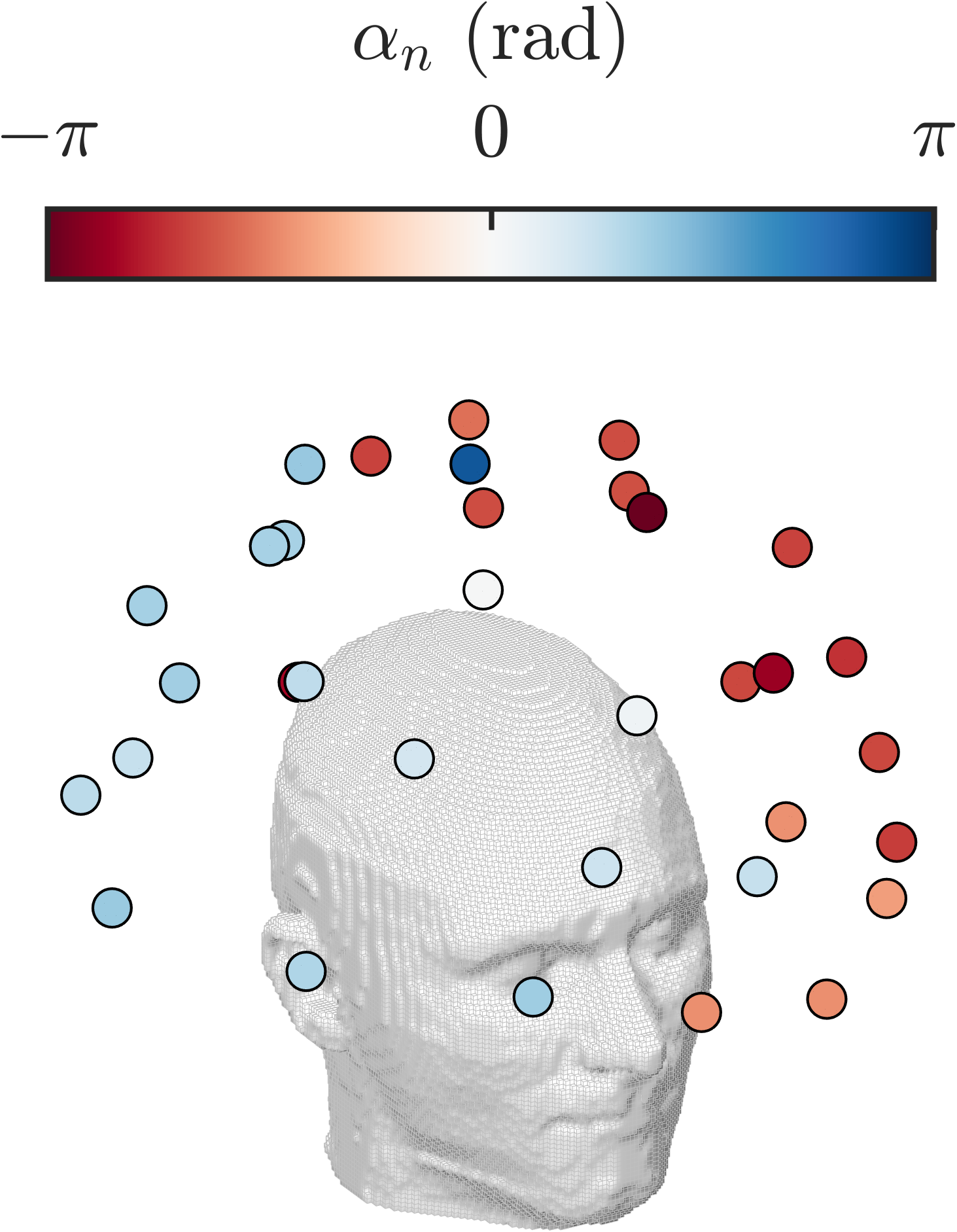}
    \label{fig:res_TI_array_a}}
    \hfil
    \subfloat[]{\includegraphics[width=0.18\textwidth]{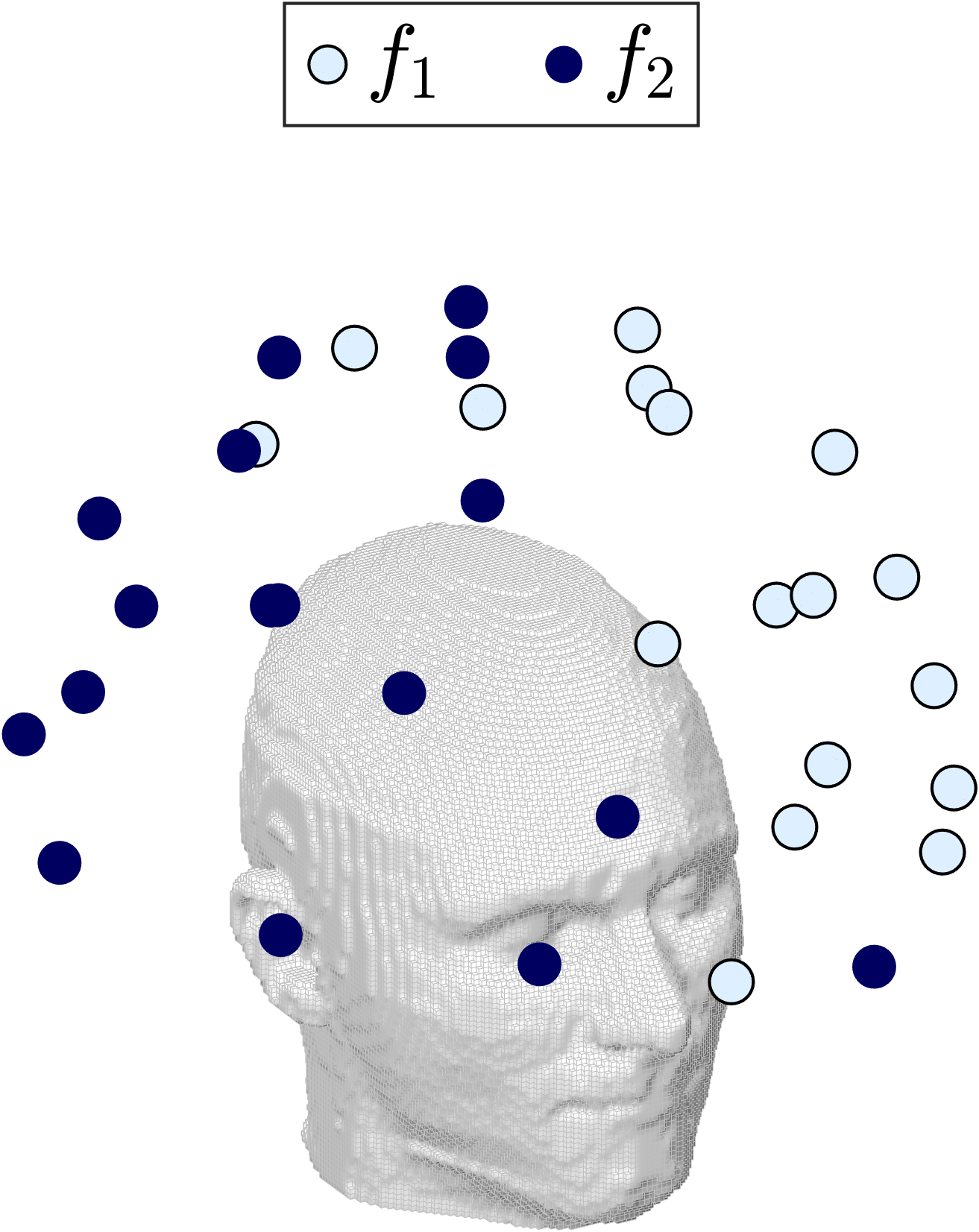}
    \label{fig:res_TI_array_s}}
    \hfil
    \subfloat[]{\includegraphics[width=0.18\textwidth]{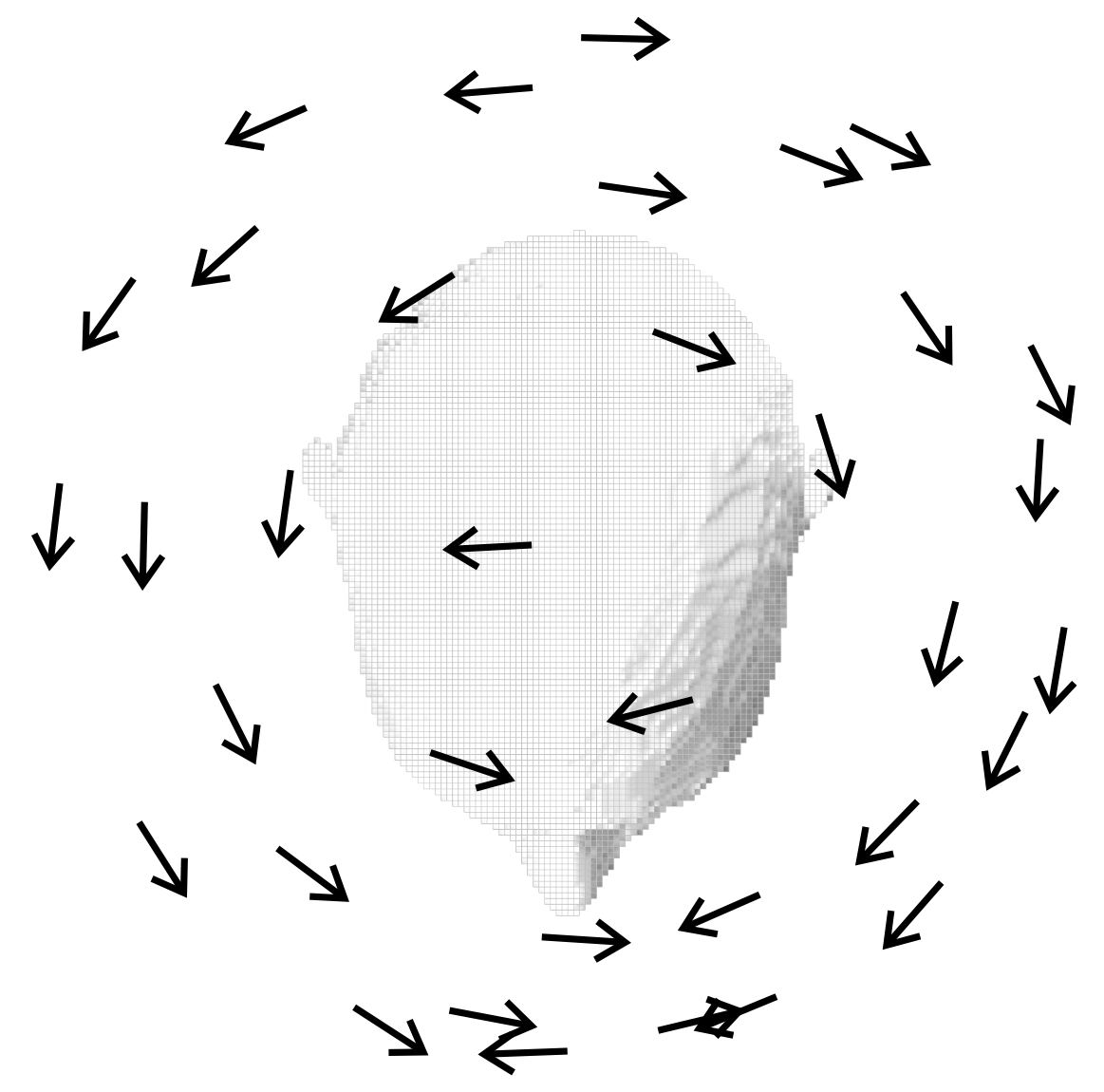}
    \label{fig:res_TI_array_d}}
    
    \caption{Antenna array properties after TI optimization for the benchmark case showing (a) dipole moment magnitudes $m_n$ (\si{\micro\ampere \meter^2}), (b) excitation phases $\alpha_n$ (rad), (c) binary frequency assignment of each element to $f_1$ or $f_2$, and (d) dipole moment orientations $\mhat_n$, shown as vectors.}
    \label{fig:res_TI_array}
\end{figure*}

The antenna array properties after TI optimization, including magnitude, phase, frequency assignment, and dipole orientation, are shown in Fig. \ref{fig:res_TI_array}. Here, the frequency assignment is split evenly between the two operating frequencies, $f_1$ and $f_2$, exhibiting approximately symmetric distribution. The dipole moment orientations are primarily confined to the $xy$-plane and follow directions approximately tangential to the head.

\begin{figure*}[ht!]
    \centering
    \includegraphics[width=0.95\textwidth]{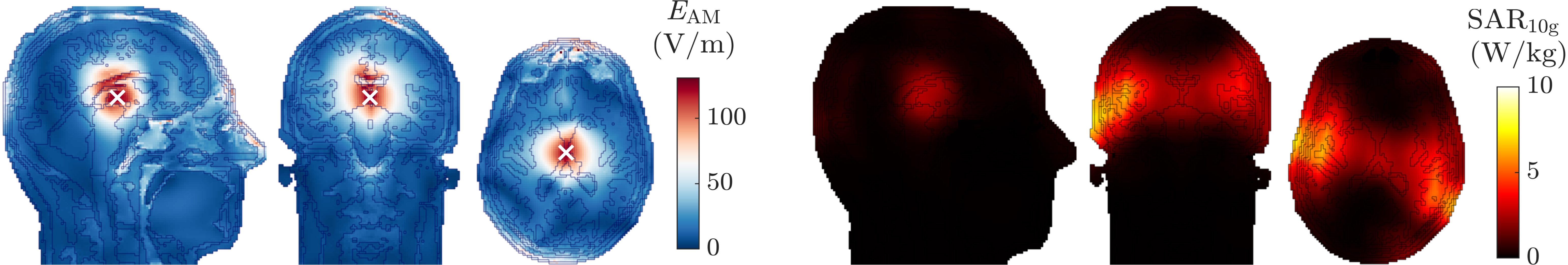}
    \caption{Distributions of $\EAM$ (V/m) (left) and $\SAR$ (W/kg) (right), for the benchmark case. The white cross indicates the defined focal point $\rfvec$.}
    \label{fig:res_base}
\end{figure*}

The results of the AM field $\EAM$ and $\SAR$ distribution after TI optimization are shown in Fig. \ref{fig:res_base}. It can be observed that the AM field $\EAM$ is confined around the center target point and $\SAR$ is below 10~W/kg, as intended.

\begin{table}[b!]
\renewcommand{\arraystretch}{1.3}
    \centering
        \caption{Benchmark configuration results over seven independent runs. The run chosen to represent the benchmark configuration is highlighted in bold shading.}
    \label{tab:res_base}
    \begin{tabular}{|c|c|c|c|c|c|c|c|c|}
    \hline
    
        Run  & 
        $F$  & 
        \multicolumn{2}{c}{$\EAMmax$ } &
        \multicolumn{3}{|c|}{$\SAR$} & 
        \multicolumn{2}{c|}{$m$ }
        
        \\
         &  (cm) & \multicolumn{2}{c}{(V/m)} & \multicolumn{3}{|c|}{(W/kg) }& \multicolumn{2}{c|}{(\si{\micro\ampere\meter^2})} 
         \\
        & & 
        \multicolumn{1}{c}{$\rvec_F$} &
        \multicolumn{1}{c}{$\rvec_\mathrm{B}$} &
        \multicolumn{1}{|c}{Skin} &
        \multicolumn{1}{c}{CSF}&
        \multicolumn{1}{c|}{Skull}
        &  \multicolumn{1}{c}{min} & \multicolumn{1}{c|}{max}
        
        \\
        \hline \hline
        1 & 3.78 & 156 & 79 & 10.0 & 9.61 & 9.10 & 5.0 & 210.6 \\ 
        2 &  3.33 & 121 & 60 & 10.0 & 8.27 & 9.44 & 4.8 & 213.5 \\ 
        3 &  3.39 & 124 & 62 & 10.0 & 7.02 & 8.25 & 0.6 & 204.9 \\ 
        4 & 3.97 & 127 & 64 & 10.0 & 8.31 & 9.00 & 9.0 & 196.5 \\ 
        5 & 3.82 & 143 & 72 & 10.0 & 8.09 & 8.98 & 2.8 & 218.6 \\ 
        \rowcolor{gray!15} \textbf{6} & \textbf{3.69} & \textbf{131} & \textbf{66} & \textbf{10.0} & \textbf{6.36} & \textbf{7.49} & \textbf{6.3} & \textbf{178.8}  \\ 
        7 & 3.88 & 127 & 64 & 10.0 & 7.76 & 7.92 & 3.5 & 175.3 \\ 
        \hline
    \end{tabular}
\end{table}

\subsection{Focal Region Radius}
To evaluate the role of the focal region definition $\Omega_\mathrm{f}$, see section \ref{sec:method focal region}, and to verify that the results are not sensitive to this numerical support parameter, the focal radius is varied over the range 5–25~mm, in steps of 5~mm. The location of the confined AM field, as well as the quantitative measures of focality and focal intensity, remained consistent across this range, showing that the framework is not sensitive to this parameter. This confirms that the focal region primarily acts as a numerical support parameter in the objective function, to localize the envelope around the target point, rather than a factor influencing focusing performance. 

\subsection{Frequency}
The influence of operating frequency on the focusing performance is evaluated. Since frequency determines both field penetration and spatial confinement in biological tissue, it directly impacts the achievable focality and absorption. The frequency is set to $f_1 = X$~Hz and $f_2=X+100$~Hz, where $X\in\{400,500,600,700,800,900,1000\}$~MHz, while all other parameters are fixed to the benchmark case.

\begin{figure*}[!t]
    \centering
    \subfloat[400~MHz]{\includegraphics[width=0.95\textwidth]{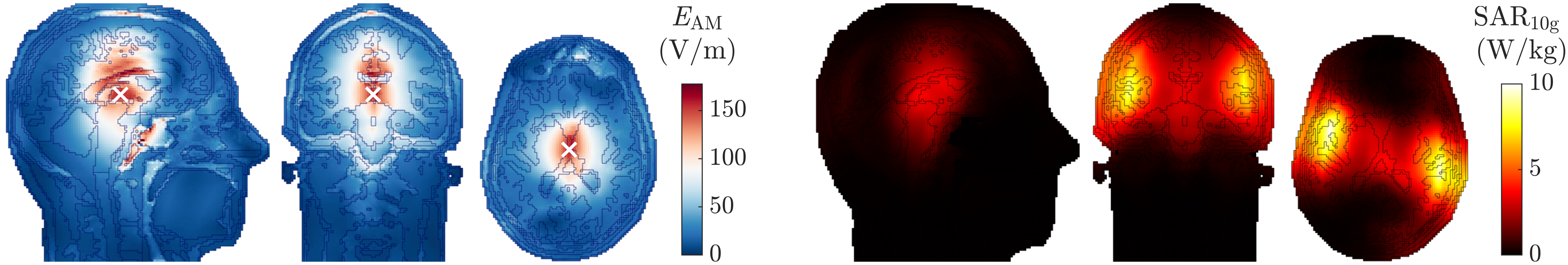}
    \label{fig:res_freq_400}}

    \subfloat[700~MHz (benchmark)]{\includegraphics[width=0.95\textwidth]{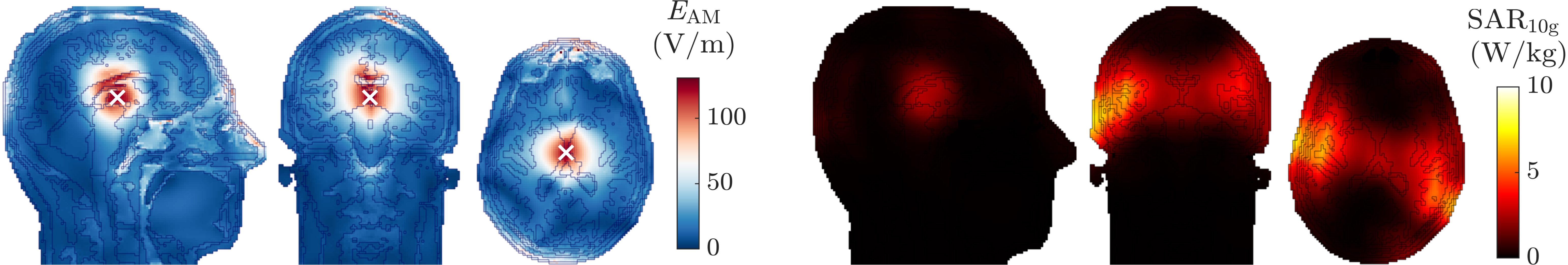}
    \label{fig:res_700}}
   
    \subfloat[1000~MHz]{\includegraphics[width=0.95\textwidth]{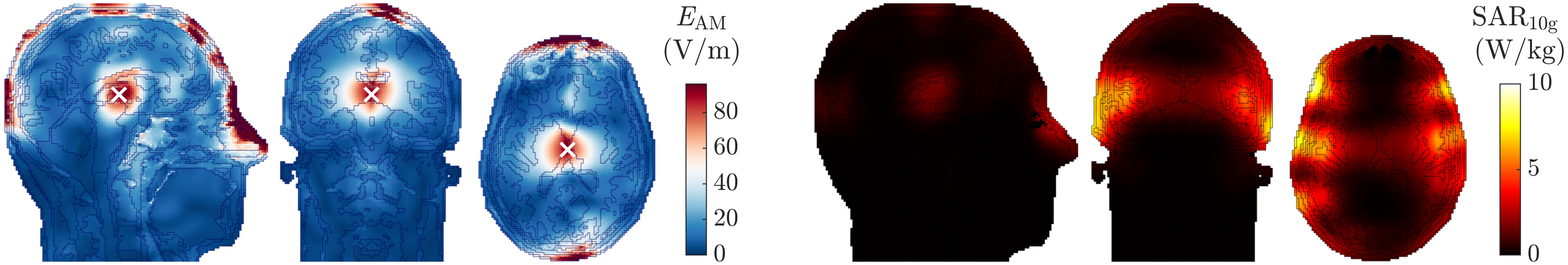}
    \label{fig:res_freq_1000}}

    \caption{Distribution of $\EAM$ (V/m) (left column) and $\SAR$ (W/kg) (right column), for different operating frequencies, with (a) 400 MHz, (b) 700~MHz (benchmark), and (c) 1000~MHz, while all other parameters are fixed to the benchmark configuration. The white cross indicates the defined focal point $\rfvec$.}
    \label{fig:res_freq}
\end{figure*}

The effect of operating frequency is summarized in Table~\ref{tab:frequency} and illustrative cases for 400~MHz, 700~MHz, and 1~GHz are presented in Fig.~\ref{fig:res_freq}. Lower frequencies provide higher intensity due to reduced attenuation, but at the cost of reduced focality. As the frequency increases, the fields become more spatially localized with improved focality, although with a reduced intensity due to increased attenuation. $\SAR$ remains below 10~W/kg for all cases.

\begin{table}[t!]
    \caption{Results for varying operating frequency, with all other parameters fixed to the benchmark configuration (benchmark row in bold shading).}
    \label{tab:frequency}
    \renewcommand{\arraystretch}{1.3}
    \centering
    \begin{tabular}{|c|c|c|c|c|c|c|c|c|}
    \hline
    
        $f$  & 
        $F$  & 
        \multicolumn{2}{c}{$\EAMmax$ } &
        \multicolumn{3}{|c|}{$\SAR$} & 
        \multicolumn{2}{c|}{$m$ }
        
        \\
        (MHz) & (cm) & \multicolumn{2}{c}{(V/m)} & \multicolumn{3}{|c|}{(W/kg) }& \multicolumn{2}{c|}{(\si{\micro\ampere\meter^2})}\\
        & &
        \multicolumn{1}{c}{$\rvec_F$} &
        \multicolumn{1}{c}{$\rvec_\mathrm{B}$} &
        
        \multicolumn{1}{|c}{Skin} &
        \multicolumn{1}{c}{CSF}&
        \multicolumn{1}{c|}{Skull}
        &  \multicolumn{1}{c}{min} & \multicolumn{1}{c|}{max}
        
        \\
        \hline \hline
        400 &  4.16 & 178 & 89 & 7.70 & 9.84 & 7.67 &  11.3 & 568.6\\ 
        500 &  3.61 & 160 & 81 & 10.0 & 9.34 & 9.55 &  14.7 & 350.9 \\ 
        600 &  3.88 & 160 & 83 & 8.09 & 9.61 & 7.93 &  17.2 & 269.8 \\ 
        \rowcolor{gray!15} \textbf{700} & \textbf{3.69} & \textbf{131} & \textbf{66} & \textbf{10.0} & \textbf{6.36} & \textbf{7.49} & \textbf{6.3} & \textbf{178.8} \\ 
        800 &  3.24 & 125 & 62 & 10.0 & 9.72 & 8.54 &  2.1 & 189.9 \\ 
        900 &  3.02 & 103 & 53 & 10.0 & 7.80 & 9.21 & 1.5 &136.1\\ 
        1000 &  3.21 & 96 & 61 & 10.0 & 6.42 & 7.34 & 1.5 & 98.9 \\ 
        \hline
    \end{tabular}
\end{table}

\subsection{Number of Antenna Elements}
The array size determines the available spatial degrees of freedom for field shaping, and is therefore expected to influence the control of shaping the AM field in the heterogeneous head. To assess this, the number of antennas is varied between $N=8,\,16,\,24$ and $35$ while all other parameters are fixed to the benchmark configuration. 

\begin{figure*}[!t]
    \centering
    \subfloat[$N=8$]{\includegraphics[width=0.95\textwidth]{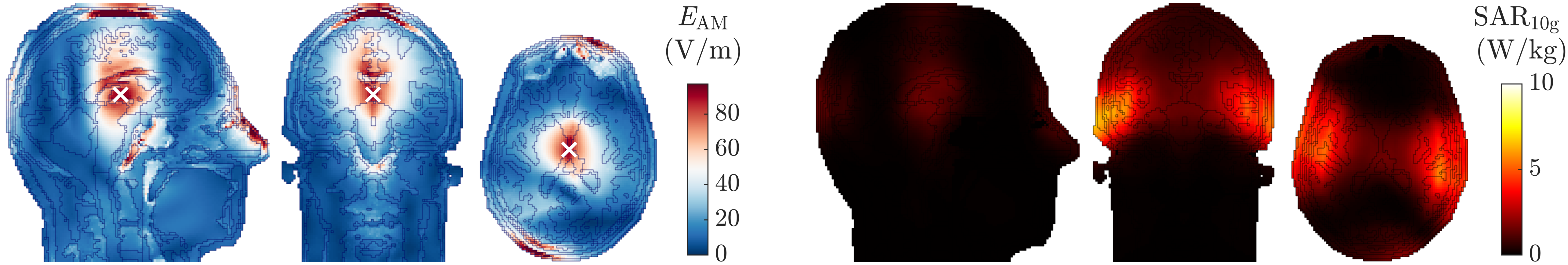}
    \label{fig:res_N_8}}

    \subfloat[$N=16$]{\includegraphics[width=0.95\textwidth]{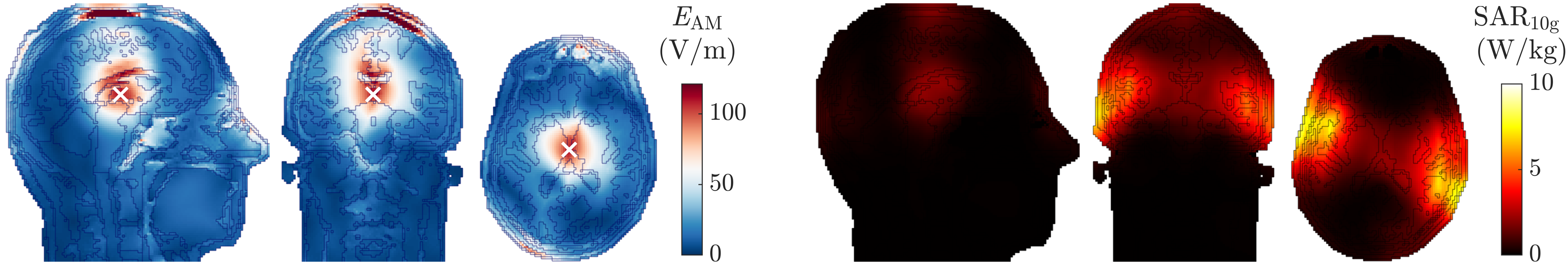}
    \label{fig:res_16}}
   
    \subfloat[$N=24$]{\includegraphics[width=0.95\textwidth]{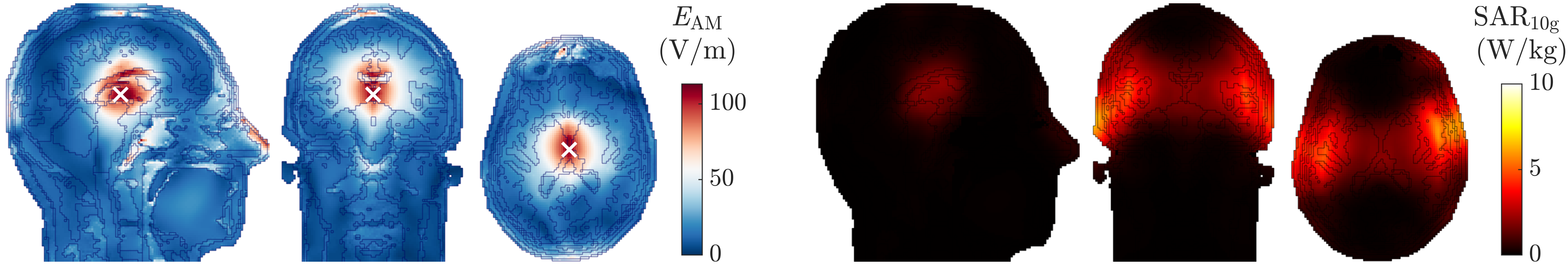}
    \label{fig:res_24}}

    \subfloat[$N=35$ (benchmark)]{\includegraphics[width=0.95\textwidth]{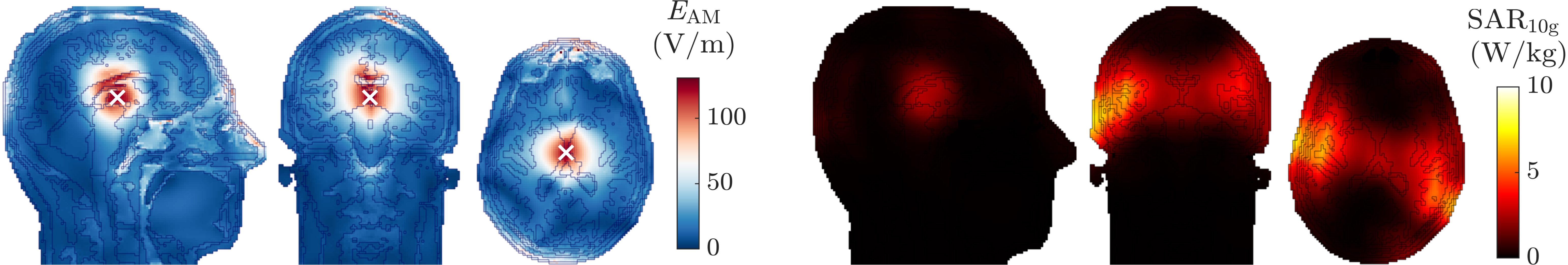}
    \label{fig:res_N_35}}

    \caption{Distribution of $\EAM$ (V/m) (left column) and $\SAR$ (W/kg) (right column), for different antenna element counts, with (a) $N=8$, (b) $N=16$, (c) $N=24$, and (d) $N=35$ (benchmark). The white cross indicates the defined focal point $\rfvec$.}
    \label{fig:res_N}
\end{figure*}

The distributions of $\EAM$ and $\SAR$ are presented in Fig. \ref{fig:res_N}. The quantitative results of focality, intensity, and $\SAR$ are presented in Table \ref{tab:N}. Increasing the number of antenna elements improves both focality and intensity due to increased spatial degrees of freedom. Smaller arrays exhibit broader fields and reduced intensity. All configurations satisfy the $\SAR$ constraint.

\begin{table}[t!]
\renewcommand{\arraystretch}{1.3}
    \centering
    \caption{Results for varying the antenna element count $N$, with all other parameters fixed to the benchmark configuration (benchmark row in bold shading).
    }
    \label{tab:N}
    \begin{tabular}{|c|c|c|c|c|c|c|c|c|c|}
    \hline
    
        $N$  & 
        $F$  & 
        \multicolumn{2}{c}{$\EAMmax$ } &
        \multicolumn{3}{|c|}{$\SAR$} & 
        \multicolumn{2}{c|}{$m$ }
        
        \\
         & (cm) & \multicolumn{2}{c}{(V/m)} & \multicolumn{3}{|c|}{(W/kg) }& \multicolumn{2}{c|}{(\si{\micro\ampere\meter^2})} 
         \\
        & & 
        \multicolumn{1}{c}{$\rvec_F$} &
        \multicolumn{1}{c}{$\rvec_\mathrm{B}$} &
        
        \multicolumn{1}{|c}{Skin} &
        \multicolumn{1}{c}{CSF}&
        \multicolumn{1}{c|}{Skull}
        &  \multicolumn{1}{c}{min} & \multicolumn{1}{c|}{max}
        
        \\
        \hline \hline
        8 & 4.57  & 97 & 50 & 10.0 & 6.00 & 6.47 & 9.1& 284.7\\ 
        16 &  4.04 & 121 & 63 & 10.0 & 7.31 &7.58 & 7.9& 230.2\\ 
        24 &  4.05 &  113 & 57 & 10.0 & 6.54 & 6.75 & 2.8& 230.6\\ 
        \rowcolor{gray!15} \textbf{35}  & \textbf{3.69} & \textbf{131} & \textbf{66} & \textbf{10.0} & \textbf{6.36} & \textbf{7.49} & \textbf{6.3} & \textbf{178.8} \\ 
        \hline
    \end{tabular}
\end{table}

\subsection{Target Locations}
To assess sensitivity to anatomical heterogeneity and target location, we perform a systematic evaluation across multiple target positions, spanning both deep and displaced regions within the brain. The center focus (0,0,0) corresponds approximately to the subthalamic nucleus (STN), a common target in invasive DBS. Additional target locations are obtained by applying spatial offsets of $\pm 2$~cm along the Cartesian directions relative to this center. Lateral displacement (+2~cm in $x$), anterior and posterior displacements (+2~cm and $-2$~cm in $y$), and superior and inferior displacements (+2~cm and $-2$~cm in $z$) are considered. These shifted locations approximate positions within regions associated with targets such as the globus pallidus internus (GPi), ventral striatum (VS), thalamus, and pedunculopontine nucleus (PPN). These locations are used to provide representative coverage, guided by clinically relevant DBS target regions \cite{Harmsen2020}.

\begin{figure*}[!t]
    \centering
    \subfloat[Center focal point $\rfvec=(0,0,0)$ cm (benchmark)]{\includegraphics[width=0.95\textwidth]{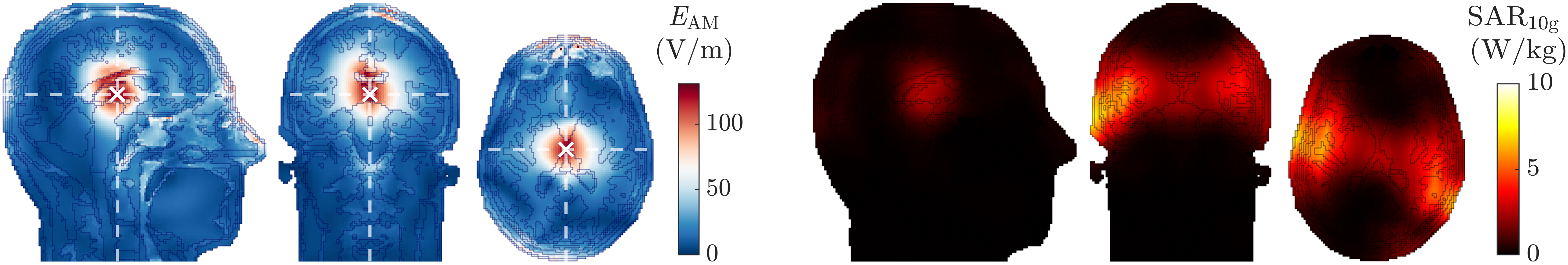}
    \label{fig:res_target_a}}

    \subfloat[Lateral focal point $\rfvec=(+2,0,0)$ cm]{\includegraphics[width=0.95\textwidth]{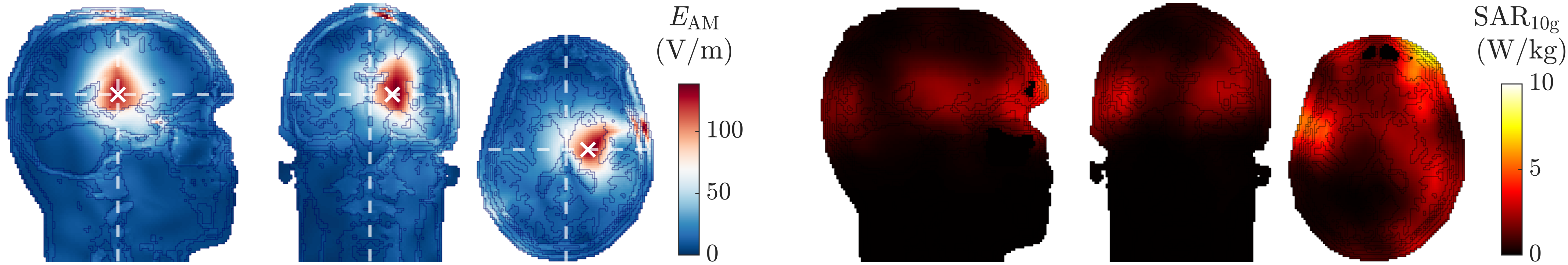}
    \label{fig:res_target_b}}
   
    \subfloat[Anterior focal point $\rfvec=(0,+2,0)$ cm]{\includegraphics[width=0.95\textwidth]{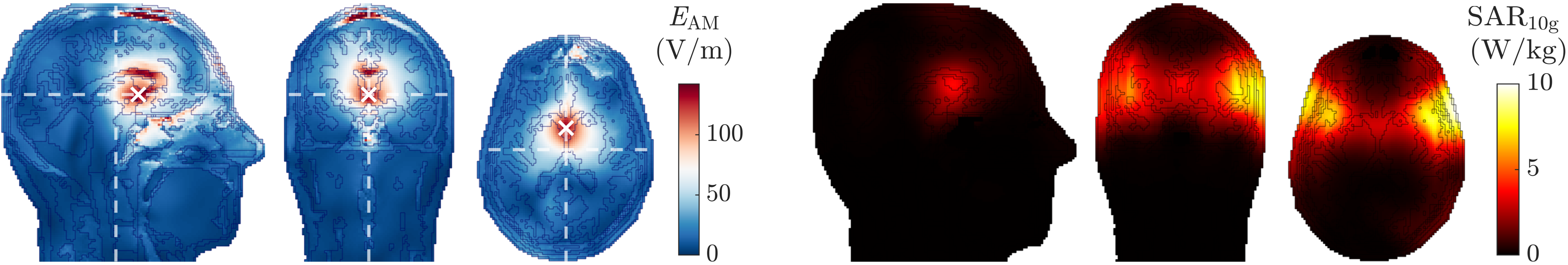}
    \label{fig:res_target_c}}

    \subfloat[Superior focal point $\rfvec=(0,0,+2)$ cm]{\includegraphics[width=0.95\textwidth]{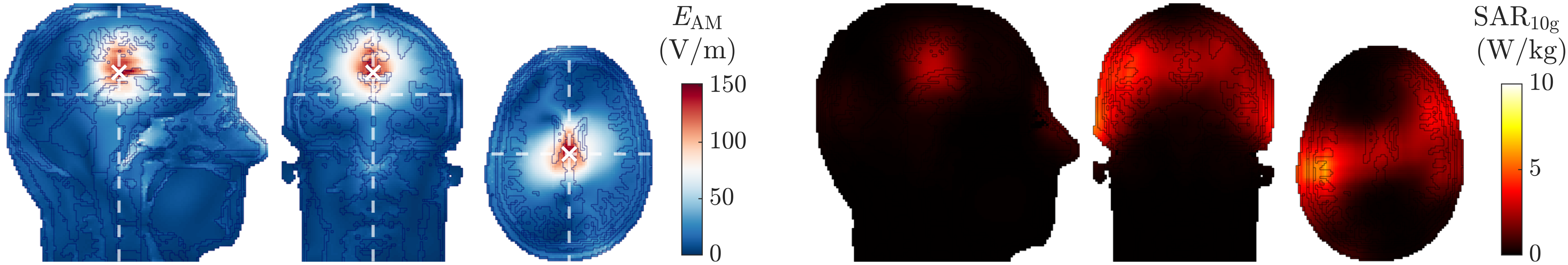}
    \label{fig:res_target_d}}

    \subfloat[Posterior focal point $\rfvec=(0,-2,0)$ cm]{\includegraphics[width=0.95\textwidth]{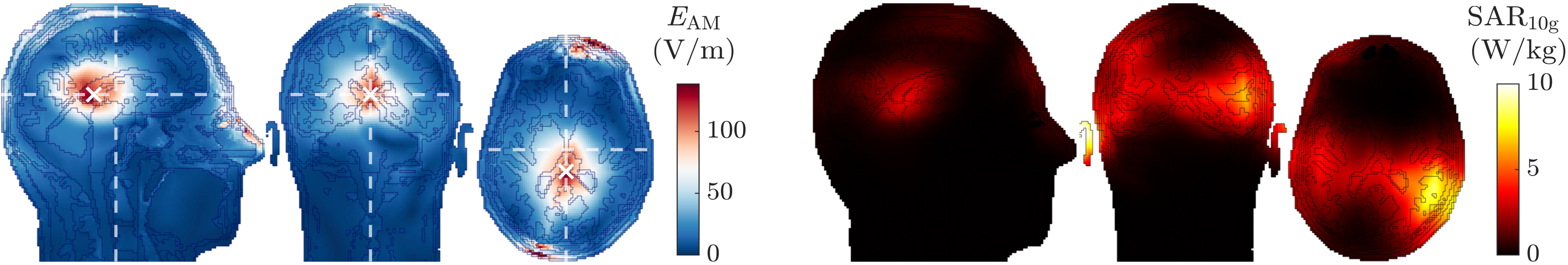}
    \label{fig:res_target_e}}

    \subfloat[Inferior focal point $\rfvec=(0,0,-2)$ cm]{\includegraphics[width=0.95\textwidth]{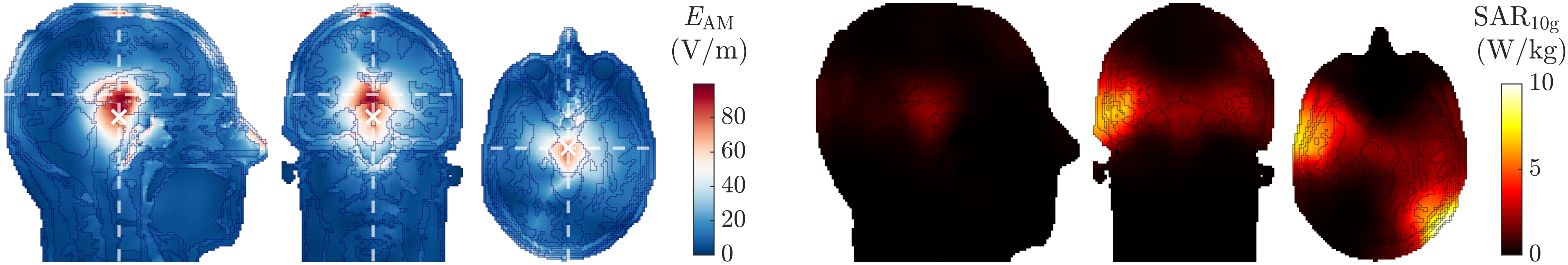}
    \label{fig:res_target_f}}

    \caption{Distribution of $\EAM$ (V/m) (left column) and $\SAR$ (W/kg) (right column), for different target locations $\rfvec$, with (a)–(f) corresponding to spatial offsets from the head center. All other parameters are fixed to the benchmark configuration. The horizontal and vertical dashed lines represent the center target location, to show the offset of each target point relative to the center target of the benchmark case. The white cross indicates the defined focal point $\rfvec$ in each case.}
    \label{fig:res_target}
\end{figure*}

The distributions of $\EAM$ and $\SAR$ are presented in Fig. \ref{fig:res_target}, where dashed lines intersect at the central target location, defining the reference axes, with offsets indicating the direction of target displacement. The results demonstrate that $\EAM$ remains well localized at the intended target for both central and displaced cases, with limited off-target modulation, although minor variations arise due to differences in tissue properties and propagation paths.

The quantitative results of focality, intensity, and $\SAR$ are presented in Table \ref{tab:target}. The focality and intensity remain within a relatively narrow range across all target locations, indicating stable focusing performance. Slight variations in intensity are observed depending on the displacement direction. The $\SAR$ values remain under 10~W/kg for all cases.

\begin{table}[t!]
    \renewcommand{\arraystretch}{1.3}
    \caption{Results for different focal point locations $\rfvec$, with all other parameters fixed to the benchmark configuration (benchmark row in bold shading).}
    \label{tab:target}
    \centering
    \begin{tabular}{|c|c|c|c|c|c|c|c|c|c|}
    \hline
    
        \multicolumn{1}{|c|}{$\rfvec$}   & 
        $F$  & 
        \multicolumn{2}{c}{$\EAMmax$ } &
        \multicolumn{3}{|c|}{$\SAR$} & 
        \multicolumn{2}{c|}{$m$ } 
        
        \\
         (cm)  & (cm) & \multicolumn{2}{c}{(V/m)} & \multicolumn{3}{|c|}{(W/kg) }& \multicolumn{2}{c|}{(\si{\micro\ampere\meter^2})}
         
         \\
        ($x, y, z$) & &
        \multicolumn{1}{c}{$\rvec_F$} &
        \multicolumn{1}{c}{$\rvec_\mathrm{B}$} &
        
        \multicolumn{1}{|c}{Skin} &
        \multicolumn{1}{c}{CSF}&
        \multicolumn{1}{c|}{Skull}
        &  \multicolumn{1}{c}{min} & \multicolumn{1}{c|}{max}
        
        \\
        \hline \hline
        \rowcolor{gray!15} \textbf{(0,0,0)} & \textbf{3.69} & \textbf{131} & \textbf{66} & \textbf{10.0} & \textbf{6.36} & \textbf{7.49} & \textbf{6.3} & \textbf{178.8}\\ 
        (+2,0,0)     & 3.98 & 138  & 79 & 9.71 &  5.27 &8.10 & 1.3 & 157.8\\ 
        (0,+2,0)     & 3.59 & 142 & 71 & 10.0 & 8.46 & 9.08 & 2.4 & 215.1 \\ 
        (0,0,+2)     & 3.46 & 152 & 76 & 10.0 & 5.66 & 5.51 & 6.7 & 175.9 \\ 
        (0,-2,0)   &  3.65 & 139 & 69 & 10.0 & 8.45 & 7.18 & 0.7 & 200.9  \\
        (0,0,-2) & 3.57 & 100 & 51 & 10.0 & 6.85 & 8.23 & 4.3 & 162.4
        \\
        \hline
    \end{tabular}
\end{table}

\subsection{Perturbation Study}
To assess the reliability of the proposed method, a perturbation study is conducted. The effects of variations in element positions, orientations, and excitation parameters are evaluated, providing insight into the stability of the focusing performance on the benchmark configuration, presented in section \ref{sec:res_base}.

Firstly, each element is displaced by a uniformly distributed random offset in all three spatial directions, with a maximum deviation of  $\pm 2$~mm. The element is then slightly tilted from its original direction by a small random angle, up to \SI{2}{\degree}.

Secondly, each element is perturbed in magnitude $m_n$ by a uniformly distributed deviation within $\pm 1.5$~\si{\micro\ampere \meter^2}, which corresponds to approximately $\pm\SI{2}{\percent}$ of the mean antenna magnitude. Each element is then perturbed in phase $\alpha_n$ by a uniformly distributed variation within \SI{2}{\percent} of $\pi$ radians.

The resulting field distributions were visually nearly identical to the reference configuration. Quantitatively, only minor deviations were observed. For perturbations in element placement and orientation, the resulting changes were -0.53\% in focality, -2.24\% in focal intensity, and a peak $\SAR$ of 10.21~W/kg. Similarly, for perturbations in magnitudes and phases, the resulting changes were +0.26\% in focality, +0.02\% in focal intensity, and a peak $\SAR$ of 10.73~W/kg. All percentage changes are reported relative to the benchmark configuration.

\begin{table*}[t!]
    \renewcommand{\arraystretch}{1.3}
    \caption{Statistical summary of all simulated configurations across the considered parameter variations.}
    \label{tab:statistical}
    \centering
    \begin{tabular}{|c|c|c|c|c|c|}
    \hline
    Metric & Mean (95\% CI) & Median & 5th-95th percentiles & Min & Max
    \\ \hline \hline
    $F$ (cm) & 3.72 (3.58--3.85)& 3.78  & 3.18--4.22 & 3.02 & 4.57 \\
    $\EAMmax(\rvec\in\Vfoc) $ (V/m) & 131 (122--139) & 127 & 97--163 &  96 & 178 \\

    \hline
    \end{tabular}

\end{table*}

\subsection{Statistical Summary}
To provide an overall assessment of the framework performance, a statistical summary is presented in Table \ref{tab:statistical}. The results are aggregated over the full set of simulations, including variations in target location, focal region radius, operating frequency, and number of antenna elements.

The table reports the mean values with corresponding 95\% confidence intervals (CI), as well as median, percentile ranges (5th–95th), and minimum and maximum values for the key performance metrics. These include the focality $F$ and intensity $\EAMmax$ within the focality volume. The distributions exhibit a relatively narrow spread across the evaluated scenarios, as shown in Table \ref{tab:statistical}.

\section{Discussion}
This section interprets the results of the proposed MW iTR and TI focusing framework in a broader physical and methodological context. It begins with the interaction between the iTR and TI stages, followed by a physical interpretation of the field behavior. The influence of key parameters, robustness, model assumptions and limitations, and medical considerations are then discussed, before concluding with implications and future work.

\subsection{Interpretation of the Proposed Framework}
The iTR stage determines the positions and orientations of the antenna elements in the array, and provides initial values for the phase and magnitude used in the subsequent TI optimization. It is based on time-reversed propagation from the target and therefore captures the relevant propagation effects of heterogeneous tissue in the array configuration. 

The TI optimization refines the solution obtained from iTR toward selective neuron stimulation, where the problem shifts to the AM field formulation. This requires a balance between field amplitudes to maximize modulation depth, while preserving the spatial confinement established by the iTR stage. The TI optimization produces a confined field at the intended target location for the considered cases, with minimal amplitude elsewhere and only limited field build-up near the outer head regions.

While iTR and TI are conceptually distinct methods, their combination is essential for achieving robust and effective performance of the proposed approach. Without iTR initialization, the TI optimization must operate over a larger set of optimization variables without a physically informed starting point, leading to slower convergence. In contrast, iTR alone cannot achieve neural stimulation, since this requires AM fields, while iTR is based on a single-frequency formulation. The two-stage approach therefore yields a more structured optimization process, enabling a natural separation between array configuration and interference optimization that supports both efficient convergence and selective neural stimulation.

\subsection{Physical Interpretation of Field Behavior}
The AM field is maximized in regions where the two fields $\E_{f_1}$ and $\E_{f_2}$ are of comparable magnitude, as reflected in the definition of $\EAM$ in Eq. (\ref{eq:EAM}). In contrast, if one field dominates while the other is weak, the resulting modulation depth is reduced and no distinct envelope is formed. Thus, field intensity alone is insufficient, and balanced field contributions are required for effective modulation. For center-focused configurations, the AM field $\EAM$ exhibits a strong peak at the target location, as intended. However, elevated field levels are also observed in specific superficial regions, most notably near the nose and the top of the head, see e.g. Figs. \ref{fig:res_freq} and \ref{fig:res_N}. 

The elevated fields in specific superficial regions can be understood by considering both the symmetry of the head model and the corresponding antenna configuration. For center-focus cases, the TI optimization yields an approximately symmetric frequency allocation. As seen in Fig. \ref{fig:res_TI_array_s} for the benchmark case, antennas on opposing sides of the sagittal ($yz$) plane are assigned to $f_1$ and $f_2$ in a balanced manner. As a consequence, regions located along this plane (e.g. the nose and top of the head) are exposed to fields originating from antennas operating at both frequencies with potentially comparable amplitudes. This creates conditions where the individual fields $\E_{f_1}$ and $\E_{f_2}$ can be of similar magnitude, which in turn leads to locally elevated values of the AM field $\EAM$. This effect is particularly evident in configurations with fewer antennas, as seen in Fig. \ref{fig:res_N_8}, where the spatial control of the fields is more limited and such overlap becomes more pronounced.

However, the optimization objective explicitly seeks to suppress $\EAM$ outside the focal region. To achieve this, the algorithm redistributes the field contributions by adjusting the amplitudes of the antenna elements across the array. This reduces the relative balance between $\E_{f_1}$ and $\E_{f_2}$ in regions of strong overlap near the symmetry plane, which suppresses $\EAM$ there. At the same time, antennas located further apart, particularly laterally near the ears, must operate at higher amplitudes to ensure sufficient field contribution to the central target. This leads to increased local electric fields and, consequently, higher SAR levels in these lateral regions. This is consistently observed in the $\SAR$ field distributions for center target configurations. This trend is also reflected in the quantitative results presented in Tables \ref{tab:frequency}--\ref{tab:target}. All cases reporting the maximum value of 10~W/kg in the skin correspond to the global peak $\SAR$ within the model and are consistently located in the ear region.

For off-center targets, the antenna frequency allocation becomes noticeably asymmetric. While elements operating at $f_1$ and $f_2$ remain separated across opposite sides of the head, the distribution is no longer balanced. Instead, a larger number of elements at one frequency is allocated to the side opposite the target displacement, compensating for longer propagation paths. This behavior was consistently observed across all off-center cases. This asymmetry enhances the modulation depth at the target while suppressing unintended amplitude modulation elsewhere, consistent with compensating for the geometrical and electrical asymmetry introduced by the shifted focal point. Furthermore, the regions of elevated superficial fields in the skin and skull are also displaced relative to the central axis (e.g., Fig. \ref{fig:res_target_c}), providing additional physical support for the observed asymmetric field distribution.

It is important to note that the goal of TI-based DBS is to selectively stimulate neurons within a target brain region, rather than to induce tissue heating as in hyperthermia. Consequently, elevated AM fields in regions such as skin, skull, or CSF, are not of concern, as no neurons are affected.

The trade-off between attenuation, field intensity, and focality arises from wave propagation in lossy tissue. Achieving high intensity in deep brain tissue requires sufficient field penetration, while strong focality depends on effective interference between the two fields, which in TI requires comparable field strengths. Unequal attenuation along different propagation paths reduces the achievable modulation depth, even when the total field remains high. At the same time, increasing field intensity is constrained by SAR limits, which restrict the total power that can be delivered. Consequently, the achievable solution reflects a balance between maximizing modulation depth at the target and maintaining safe energy absorption levels throughout the head, resulting in a fundamental trade-off that reflects the underlying physics rather than limitations of the optimization approach.

\subsection{Influence of Key Parameters}
\subsubsection{Operating Frequency Effects}
Frequency governs the trade-off between field penetration and spatial confinement in biological tissues. At lower frequencies, reduced attenuation allows the EM fields to penetrate deeper into the head model, resulting in higher achievable intensity within the focal region. This increased penetration, however, is accompanied by a higher focality value, i.e., reduced spatial confinement due to the longer wavelength. At higher frequencies, the fields become more localized due to the shorter wavelength, improving confinement, but with a corresponding reduction in focal intensity. This reduction is primarily due to increased dielectric losses in tissue and the associated SAR constraints. These results demonstrate a clear frequency dependency. Consequently, the choice of operating frequency becomes a key design parameter, requiring careful selection depending on whether the primary objective is to maximize focal intensity or to achieve high spatial precision in the targeted stimulation location.

\subsubsection{Array Size Effects}
The number of antenna elements in the array determines the available spatial degrees of freedom for shaping the AM field. Increasing the array size enables more precise control over both constructive interference at the target location and destructive interference in surrounding regions, which is reflected in the observed improvements in both focality and achievable intensity seen in Table \ref{tab:N}. In contrast, reducing the number of antennas limits this control, resulting in a broader focal region, lower intensity in the focal volume, and higher intensity in superficial regions such as skin and skull. The results show clear performance gains with increasing array size. However, the improvement is subject to practical constraints, including increased system complexity and limitations in physically realizable array configurations. A lower number of antennas may be sufficient to achieve the required level of neural stimulation, while reducing the complexity of the array. Thus, the number of antennas represents an important system parameter that influences the balance between focusing performance and practical feasibility.

\subsubsection{Target Location Dependence}
The influence of target location provides insight into the robustness of the proposed framework in the presence of anatomical heterogeneity. The results demonstrate that consistent and localized focusing can be achieved across all considered target positions, as summarized in Table \ref{tab:target}. This indicates that the combined iTR-TI optimization effectively compensates for variations in propagation paths and tissue properties, adapting the antenna excitations to maintain the desired interference pattern. Although the focusing performance remains stable, minor variations in focality and intensity are observed depending on the target location. It is observed that the field intensity increases as the target location approaches the superficial regions. As shown in Table \ref{tab:target}, the point $(0,0,-2)$, which lies deeper within the head than the central reference target, exhibits a lower intensity, whereas the point $(0,0,+2)$, located closer to the superficial layers, yields the highest observed intensity. This behavior is due to shorter propagation paths and reduced attenuation for superficial targets compared to deeper targets. It can be observed that the field exhibits localized build-up in certain anatomical regions of the brain (e.g., Fig. \ref{fig:res_target_f}). This variation is due to local differences in dielectric properties and geometric asymmetries within the head model. Nevertheless, these results indicate that the proposed framework is capable of achieving consistent and controlled focusing across anatomically distinct regions relevant for DBS, demonstrating its robustness to variations in target location.

\subsection{Robustness and Stability}
The robustness and stability of the proposed framework is assessed by analyzing its sensitivity to perturbations in antenna parameters and by evaluating statistical variations across multiple configurations.

For the perturbation study, only minor changes are observed in focality and intensity, indicating that the focusing performance is not overly sensitive to small perturbations. However, in both cases, the maximum $\SAR$ exceeds the limit of 10~W/kg. This shows that although the field at the target remains stable, where all antenna elements jointly contribute, the highest $\SAR$ occurs in lateral superficial regions, where fewer elements contribute to shape the field. As a result, these regions are less controlled and more sensitive to parameter deviations, making $\SAR$ more susceptible to perturbations. This highlights the need for careful consideration of safety, and implies that a more conservative $\SAR$ margin should be applied during optimization to account for imperfections in the physical realization of the array.

The statistical summary (Table \ref{tab:statistical}) shows a relatively narrow spread in the distributions. This indicates that the proposed framework yields consistent focusing performance despite variations in anatomical location and antenna array parameters.

Together, the perturbation and statistical analyses confirm that the proposed framework is both robust to practical uncertainties and stable across a broad set of operating conditions. However, the observed sensitivity of $\SAR$ to perturbations indicates that conservative safety margins should be applied to ensure compliance under realistic implementation conditions.

\subsection{Medical Safety Considerations}
As noted, elevated AM fields in superficial non-neural tissues are not of concern for stimulation, and SAR levels remain within safe limits. For MW-DBS, the safety requirement is simply to ensure that all AM fields remain below established stimulation thresholds outside the focal region, and that any resulting SAR or thermal increases remain within safety limits throughout the head. In hyperthermia, tissue heating is intentionally produced within a target region while staying below limits elsewhere. Since TI-based neuromodulation aims to avoid heating at the target, meeting these limits is feasible. As a precaution, any potential thermal effects in superficial layers can be mitigated by using, e.g. a water bolus to cool superficial tissues \cite{Drizdal2021}. 

Compared to the related MW-TI study \cite{Ahsan2022}, which employs implanted antennas inside the skull, the present approach is fully non-invasive and addresses a more clinically relevant and challenging problem. The achieved focality is comparable or slightly broader, while substantially higher focal intensity, on the order of a magnitude, is obtained under the same SAR constraints. This is likely because implanted antennas produce high local SAR that limits excitation levels, whereas the external array allows greater flexibility and avoids direct interaction with tissue. This enables safer treatment and makes antenna arrays a more attractive solution for MW-DBS. It also provides a larger margin for reaching neuromodulation thresholds and enables further optimization toward improved field confinement.

\subsection{Model Assumptions and Limitations}
\subsubsection{Idealized Source Model}
The use of MPDs enables the optimization framework to focus on field shaping, steering, and frequency allocation independently of antenna-specific design constraints. As described in Section \ref{sec:method antenna array}, the MPDs represent idealized, non-interacting sources, and therefore do not capture coupling or other antenna-dependent effects. Consequently, the reported results should therefore be interpreted as an estimate of performance under idealized conditions, providing a benchmark against which future implementations with physically realizable antenna models can be compared.

\subsubsection{Antenna Placement and Array Geometry}
In this study, candidate antenna positions were restricted to a Fibonacci sphere within the matching layer. For the center-focus scenario, the focal point is physically equidistant from all potential antenna positions, while the effective electrical path lengths differ due to variations in tissue permittivity along each trajectory. This suggests that alternative antenna configurations could improve the reconstruction of the forward field.

\subsubsection{Tissue and Neuronal Modeling Assumptions}
The objective function $\JTI$ defined in Eq. \eqref{eq:obj_TI} is used to maximize the envelope amplitude in the target region, effectively focusing the AM field. However, this formulation does not explicitly account for the orientation of neural structures. Neuronal activation thresholds are generally anisotropic with respect to the direction of the applied electric field \cite{Wu2022}, but the approach treats stimulation as an isotropic response, which may not fully capture the underlying neurophysiological behavior. In addition, the tissue properties in the model are assumed to be isotropic, whereas realistic brain structures, such as white-matter tracts, exhibit anisotropic conductivity due to the organized orientation of axons.

\subsubsection{Field-Based Evaluation Metrics}
The focality metric is defined using a threshold of 50\% of the maximum AM field, consistent with other TI studies \cite{Ahsan2022}. This provides a robust and model-independent measure of spatial confinement, enabling comparison across configurations and between independent studies. However, it does not directly correspond to a physiological stimulation threshold, as there is currently no consensus on the electric field levels required for neural activation at MW frequencies.

\subsection{Implications and Future Work}
\subsubsection{Optimization and Objective Function}
The objective functions and optimization methods for the iTR and TI stages can be further refined. Rather than confining the field around a target point, the distribution could be optimized over a defined anatomical region. The present method emphasizes spatial selectivity but does not explicitly account for field magnitude. A natural extension is a multi-objective optimization framework, such as a multi-objective genetic algorithm (MOGA) \cite{Baskaran2023}, to jointly maximize $\EAM$, enforce spatial confinement, and minimize energy absorption. These objectives may be partially conflicting, but MOGA enables the selection of solutions that balance stimulation efficacy and safety. Modulation depth may be further enhanced by enforcing balanced field contributions at the focal point, for example, by using constraints as in \cite{Ahsan2022}.

Furthermore, no fundamental limitations prevent adding additional optimization stages. As a potential improvement, intermediate steps could be introduced between iTR and TI to provide a refined initial guess of steering parameters, potentially improving convergence or field focusing. However, the proposed two-stage framework already achieves robust focusing, such that any added complexity must be justified through a systematic evaluation of the trade-off between computational cost and performance.

\subsubsection{Development of the Antenna Array}
Extension to physically realizable antenna designs is expected to further improve performance and provide a more accurate representation of practical systems. Such implementations can provide more directive radiation patterns and enable optimization with respect to frequency, array geometry, mutual coupling, and interactions with surrounding head tissues.

\subsubsection{Neurophysiological Relevance}
Future work should extend the framework toward more detailed neurophysiological representations. A natural extension is to account for nerve tract orientations, given the anisotropic nature of neural activation \cite{Wu2022}, by projecting the electric field along these directions \cite{Grossman2017} and modifying the objective $\JTI$ in Eq.~(\ref{eq:obj_TI}) accordingly. In addition, extending the EM model to include anisotropic tissue properties would provide a more realistic representation of field propagation in structured brain regions.

To further strengthen neurophysiological relevance, the computed fields could serve as input to single-neuron models to determine stimulation thresholds for TI at MW frequencies \cite{Klooster2021}. This enables replacing the fixed 50\% focality threshold with physiologically derived activation thresholds, allowing optimization that exceeds thresholds within the target region while remaining subthreshold elsewhere. This approach would allow estimation of the activated tissue volume, analogous to metrics used in invasive DBS.

Together, these extensions would advance the understanding of the stimulation efficacy and biological effects of the proposed approach for non-invasive MW-DBS.

\section{Conclusion}
In this work, we presented a computational framework for non-invasive microwave temporal interference (TI) deep brain stimulation using an external antenna array. 
By combining iterative time reversal (iTR) with genetic-algorithm based TI optimization, the framework jointly addresses electromagnetic field focusing and physiologically relevant neural stimulation in an anatomically realistic voxel head model. The use of magnetic point dipoles provides a design-independent benchmark for evaluating focusing performance independently of specific antenna implementations.
Systematic numerical investigations demonstrated that spatially localized amplitude-modulated fields can be generated across multiple target regions, while maintaining robustness to parameter variations and perturbations. Statistical analyses further confirmed the consistency of the obtained solutions. Safety was assessed through specific absorption rate (SAR) calculations, showing compliance with established exposure limits for all considered configurations. To the best of the authors' knowledge, this is the first study to combine iTR and TI optimization for microwave-based deep brain stimulation in a realistic voxel head model. The results demonstrate the feasibility of achieving localized and safe stimulation using externally applied microwave fields, while providing insight into the trade-offs between focality, penetration depth, and safety. Overall, the proposed framework establishes a foundation for future development of non-invasive microwave neuromodulation technologies. Future work will focus on physically realizable antenna designs, improved neurophysiological models, and multi-objective optimization balancing stimulation efficacy and safety.



\section*{Acknowledgment}
We acknowledge R. Nilsson and H. Dobšíček Trefná for discussions and input on the iTR implementation, and M. Lindberg and M. Lagneskog for their assistance in implementing the focality calculation.

\bibliography{sources}
\bibliographystyle{IEEEtran}

\end{document}